\documentclass[11pt,a4paper]{article}
\usepackage{jheppub}
\usepackage{amsfonts}
\usepackage{amsmath}
\usepackage{epsfig}

\def\beq{\begin{equation}}
\def\eeq{\end{equation}}
\def\bea{\begin{eqnarray}}
\def\eea{\end{eqnarray}}
\def\bq{\begin{quote}}
\def\eq{\end{quote}}

\def\simlt{\stackrel{<}{{}_\sim}}
\def\simgt{\stackrel{>}{{}_\sim}}

\title{Gravitino dark matter with constraints \\   $\phantom{a}$ \\from Higgs boson mass and sneutrino decays}

\author[a,1]{Leszek Roszkowski\note{On leave of absence from the University of Sheffield, UK.}}
\author[a]{Sebastian Trojanowski}
\author[b]{Krzysztof Turzy\'nski}
\author[c]{Karsten Jedamzik}

\affiliation[a]{National Centre for Nuclear Research, Ho\.za 69, 00-681, Warsaw, Poland}
\affiliation[b]{Institute of Theoretical Physics, Faculty of Physics, University of Warsaw, Ho\.za 69, 00-681, Warsaw, Poland}
\affiliation[c]{Laboratoire de Physique Theorique et Astroparticules, UMR5207-CRNS, Universit\'e Montpellier II, 34095 Montpellier, France}

\emailAdd{L.Roszkowski@sheffield.ac.uk}
\emailAdd{Sebastian.Trojanowski@fuw.edu.pl}
\emailAdd{Krzysztof-Jan.Turzynski@fuw.edu.pl}
\emailAdd{karsten.jedamzik@univ-montp2.fr}

\abstract{We investigate gravitino dark matter produced thermally at high temperatures and in decays of a long-lived sneutrino. We consider the Non-Universal Higgs Model and a generalized gauge mediation
model, and in each case identify sneutrino LOSP regions of the parameter space consistent with the mass of the Higgs-like boson observed at the LHC. We apply relevant collider and cosmological bounds, including constraints from Big Bang Nucleosynthesis and from warm dark matter on large scale structures. Generally, we find allowed values of the reheating temperature $T_R$ below~$10^9$~GeV, i.e.~somewhat smaller than the values needed for thermal leptogenesis, even with a conservative lower bound of 122~GeV on the Higgs mass. Requiring mass values closer to 126~GeV implies $T_R$ below~$10^7$~GeV and the gravitino mass less than ~$10$~GeV.}

\keywords{Supersymmetry Phenomenology, Cosmology of Theories beyond the SM}

\arxivnumber{1212.5587}

\begin{document}

\maketitle
\flushbottom

\tableofcontents

\section{Introduction}

Of all extensions of the Standard Model of elementary particles, its
supersymmetric version (Minimal Supersymmetric Standard Model, MSSM)
still remains particularly well motivated (for a review, see, e.g.,\
\cite{Martin:1997ns}). Several mechanisms have been proposed to
describe the origin and mediation of the necessary supersymmetry
breaking in the MSSM, yielding distinctive mass spectra for the
supersymmetric partners of the known particles.  Among these, models
equipped with $R$-parity predict that the lightest supersymmetric
particle (LSP) is stable which allows for a possibility that it
constitutes dark matter (DM) in the Universe.

In the MSSM, the most popular DM particles are: the lightest
neutralino, the gravitino (present in the MSSM embedded in
supergravity) or the axino (in the MSSM extended with a $U(1)_\mathrm
{PQ}$ symmetry) \cite{Covi:1999ty}.  Different properties of these
particles require different variants of the history of the early
Universe, including the observationally determined abundance of
DM. The lightest neutralino LSP is considered as perhaps the most
natural choice for DM as its relic abundance from thermal freeze-out
can agree with observations, $\Omega_\mathrm{CDM}h^2=0.113\pm0.004$
\cite{Komatsu:2010fb}, for a $\mathcal{O}(100)\,\mathrm{GeV}$,
although scenarios in which the lightest neutralinos are produced in
decays of heavier particles previously dominating the energy density
of the Universe, have also been proposed, see, e.g.,\
\cite{Choi:2008zq}.

The abundance of extremely weakly interacting particles (EWIMPs), like
the gravitino or the axino LSP, is determined in a way markedly
different from that of the lightest neutralino.  They are produced in
scatterings of other particles in the primordial plasma; the abundance
of such thermally produced EWIMP is proportional to the reheating
temperature $T_R$.
In addition to  thermal production (TP), EWIMPs can also be
produced in the decays of the next-to-lightest supersymmetric
particles (NLSP), which are usually the lightest ordinary
supersymmetric particles (LOSP). While in general nearly any MSSM particle can be the
LOSP, the most natural choices, other than the lightest neutralino,
are: the lighter stau (or stop), or the sneutrino. It is this last
case that we will consider in this paper.
 %, but 
%for light gravitinos this nonthermal production from LOSP's is insignificant for the total abundance.

In non-thermal production (NTP) from LOSP decays, the LOSP lifetime may be
long enough for energetic decay products to affect the abundances of
the light elements (particularly in the case of gravitinos, while much
less so for axinos). A good agreement of predictions of the standard
Big Bang Nucleosynthesis (BBN) with observations sets stringent limits
on such additional contributions
%lr  (for a review, see e.g.\ \cite{Iocco:2008va}).
%lr  There has been a remarkable amount of work aiming at setting up the
%lr  constraints that successful standard BBN imposes on such injections of
%lr  energetic particles 
(for reviews, see e.g.\
\cite{Jedamzik:2009uy,Iocco:2008va,Pospelov:2010hj}). The resulting
picture can be roughly summarized as follows: the constrains are
weakest for light (small energy release in decay), short-lived
($\tau<\mathcal{O}(100\,\mathrm{sec})$) and not too-abundant decaying
particles, but above all for decays with a small
hadronic branching ratio.  Therefore, of all possible LOSPs,
sneutrino decays are the least constrained by BBN
since hadronic showers can only be produced
in (strongly suppressed) 3- and 4-body sneutrino decays.  It should
also be noted that the present LHC data, albeit quite restrictive for
the gluinos and the colored scalar partners of quarks of the first and
second generation, still allow sleptons and sneutrinos with much lower
masses. All this makes supersymmetric models with gravitino LSP and
sneutrino LOSP quite attractive phenomenologically from the bottom-up
perspective.\footnote{The case with light sneutrino LSP, constituting even a small portion
of DM, is very strongly constrained by direct detection experiments \cite{Falk:1994es};
in fact, the results of the XENON100 experiment require this
contribution to be at most  of the order of $10^{-3}$
of a relic sneutrino abundance.}

In the simplest scenarios of supersymmetry breaking the sneutrino is
not the LOSP and there have been just a handful of studies devoted to
analyzing BBN constraints on sneutrino LOSP
\cite{Kawasaki:1994bs,Kanzaki:2006hm,Covi:2007xj,Ellis:2008as} in
theoretically motivated scenarios, such as the Non-Universal Higgs
Model (NUHM)
\cite{Polonsky:1994sr,Matalliotakis:1994ft,Olechowski:1994gm} or
the Generalized Gauge Mediation (GGM) model
\cite{Shih:2007av,Cheung:2007es,Meade:2008wd,Carpenter:2008wi}.

In the present letter we re-visit the possibility of gravitino dark
matter from both TP and from NTP of sneutrino decays in light of
recent discovery at the LHC of a Higgs-like boson with a mass of
126~GeV \cite{:2012gk,:2012gu}. 
We will show that taking this new
result at face value implies a stringent upper bound on the
reheating temperature $T_R$  below~$10^7$~GeV, and also favors a low gravitino mass
region, below ~$10$~GeV. Assuming a conservative lower bound on the Higgs boson mass of
122~GeV leads to weaker constraints on $T_R$ and allows a larger gravitino mass.

So far $T_R$ has been allowed to take the
largest values for the sneutrino LOSP of all MSSM choices owing to the lowest yield at
freeze-out. This had a double effect of having the weakest effect on
BBN and also requiring largest $T_R$ for the TP contribution to make
up for the reduced gravitino relic abundance from NTP. However, the
relatively large Higgs mass implies larger SUSY breaking mass scale,
thus also larger masses of gauginos whose scatterings dominate
TP. This in turn boost the TP contribution (which is proportional to
their square) and, as a consequence, requires lower $T_R$. A larger
SUSY breaking scale also implies larger sneutrino mass, and, as a
consequence, larger yield at freeze-out and more energetic hadronic showers, which
translates to more stringent BBN bounds.

In addition to analyzing the effect of the Higgs mass on the scenario,
we extend and update previous analyses in several directions.  First,
for small and moderate $\tan\beta$ we identify the patterns of soft
supersymmetry breaking masses at the high scale that can lead to the
sneutrino NLSP. In realistic models, sneutrinos are often degenerate
in mass with right-handed sleptons and/or neutralinos (in addition to
degeneracy with corresponding left-handed charged sleptons). This
leads to many coannihilation channels, which may affect the relic
density $\Omega_\mathrm{NLSP}h^2$ and, as a consequence, also BBN bounds, as
they are sensitive to $\Omega_\mathrm{NLSP}h^2$.  Second, we implement
the BBN bounds using a state-of-the-art numerical code for solving the
relevant Boltzmann equations \cite{Jedamzik:2006xz}. As an input
parameter to that code, we perform a full computation of the hadronic
branching fraction of sneutrino decays, including quark-antiquark
production from on-shell and off-shell electroweak gauge bosons and
gauginos.

The paper is organized as follows. In Section 2, we review the
constraints in $\tilde \nu$ LOSP scenarios, presenting the assumptions
leading to sufficiently light sneutrinos, together with the
constraints from the Higgs boson mass measurement and BBN. In Section
3, we perform a numerical analysis of $\tilde\nu$ LOSP scenarios,
discussing the impact of various constraints on the NUHM and the GGM
model. We present our conclusions in Section 4.

\section{Review of constraints in sneutrino LOSP scenarios}

\subsection{Soft supersymmetry breaking masses at the low scale}
\label{sec:snulow}

The $\tau$-sneutrino, ${\tilde \nu}_\tau$, is the lightest of the sneutrinos
due to the $\tau$-Yukawa coupling driving its mass slightly below the
sneutrinos of the 
other two generations, and from now on we will refer to it as simply
the sneutrino.  The sneutrino can become lighter than its charged slepton
partner thanks to the electroweak $D$-term contributions to the  slepton
and the sneutrino masses. For moderate $\tan\beta$ the sneutrino mass
after electroweak symmetry breaking (EWSB) reads (see e.g.\ \cite{Haber:1984rc})
%%%%%%%%%%%%%%%%
\begin{equation}
\label{eq:snumass}
m^2_{\tilde \nu}=m^2_L+D^2_{\tilde\nu} \, ,
%-\frac{1}{2}M_Z^2\, ,
\end{equation}
%%%%%%%%%%%%%%%%
whereas the mass matrix of the charged sleptons of the third
generation is given by
%%%%%%%%%%%%%%%%
\begin{equation}
\label{eq:slmat}
\mathbf{m}^2_{\tilde\tau_{L,R}} =\left( \begin{array}{cc} m_L^2
+D^2_{\tilde\ell_L}
%+M_W^2-\frac{1}{2}M_Z^2 
& m_\tau(\mu\tan\beta-A_\tau) \\
m_\tau(\mu\tan\beta-A_\tau) & m_E^2 +D^2_{\tilde\ell_R}
%+M_Z^2-M_W^2 
\end{array}\right) \, .
\end{equation} 
%%%%%%%%%%%%%%%%
In (\ref{eq:snumass}) and (\ref{eq:slmat}) $m_L^2$ and $m_E^2$ denote
soft supersymmetry breaking mass parameters of the superpartners of
the left-handed and the right-handed leptons, respectively, and $A_\tau$
stands for the $\tau$ trilinear parameter, with all the parameters
evaluated at the EWSB scale.  The $D$-term contribution to
the sneutrino mass, $D^2_{\tilde\nu}=-\frac{1}{2}M_Z^2$, is negative,
while analogous contributions to the masses of the charged sleptons,
$D^2_{\tilde\ell_L}=M_W^2-\frac{1}{2}M_Z^2$ and
$D^2_{\tilde\ell_R}=M_Z^2-M_W^2 $, are positive.  

The sneutrino mass
(\ref{eq:snumass}) is smaller than the smaller of the eigenvalues of
the slepton mass matrix (\ref{eq:slmat}) if the condition
%%%%%%%%%%%%%%%%
\begin{equation}
\label{eq:snucon}
m_E^2 - m_L^2 >\frac{m_\tau^2(\mu\tan\beta-A_\tau)^2}{M_W^2}+M_W^2-\frac{3}{2}M_Z^2\
\end{equation}
%%%%%%%%%%%%%%%%
is satisfied. For typical values of the parameters, it follows from
the inequality (\ref{eq:snucon}) that the sneutrino is the lightest
slepton if $m_L^2<m_E^2$ and the left-right mixing in the slepton
sector is not too large. For example, with $\mu=1000\,\mathrm{GeV}$
and $\tan\beta=10$, the condition (\ref{eq:snucon}) is satisfied if the
splitting between $\sqrt{m_E^2}$ and $\sqrt{m_L^2}$ is of at least about
$100\,\mathrm{GeV}$.  It is also clear that increasing $\tan\beta$ while
keeping the other parameters in (\ref{eq:snumass}) and (\ref{eq:slmat})
fixed decreases the mass of the lighter charged slepton of each generation, eventually
closing the region of the parameters  where the sneutrino is the
lightest slepton.

%%%%%%%%%%%%%%%%
\subsection{Conditions for $m_L^2<m_E^2$ from renormalization group running from a high scale}
\label{sec:snurge}
%%%%%%%%%%%%%%%%

Since sneutrino LOSP disfavors large $\tan\beta$, we can, 
following the method outlined in \cite{Carena:1996km}, obtain solutions of renormalization group equations (RGEs) for the soft supersymmetry breaking parameters $m_E^2$  and $m_L^2$:
%%%%%%%%%%%%%%%%
\begin{eqnarray}
\label{eq:solrge}
m_E^2 &=& m_{E,0}^2+c_{E1}M_1^2+\tilde c_{EU}m_{U,0}^2-\frac{1}{11}D^2
\left(1- \frac{g_1^2}{g_{1,0}^2}\right) + \delta^2_{E,y_\tau} \\ 
\label{eq:solrgea}
m_L^2 &=& m_{L,0}^2+c_{L1}M_1^2+c_{L2}M_2^2 +\tilde
c_{LQ}m_{Q,0}^2+\frac{1}{22}D^2 \left(1-
  \frac{g_1^2}{g_{1,0}^2}\right) + \delta^2_{L,y_\tau}  
\end{eqnarray}
%%%%%%%%%%%%%%%%
where by $m_{S,0}^2$ (with an additional index 0) for $S=E,L,Q,U,D$ we denote sfermion masses at the high scale, while
$M_{1,2}$ are low-scale $U(1)$ and $SU(2)$ gaugino soft mass
parameters. The coefficients $c_{E1}$ and $c_{Li}$ can be found by
solving the 1-loop RGEs, whereas  $\tilde{c}_{EUi}$, $\tilde{c}_{LQ}$ by
solving the 2-loop RGEs and identifying the leading effects;  they are given in Table
\ref{t1} for some representative choices for the high scale $Q$ and the scale
$m_S=\sqrt{m_{\tilde t_1}m_{\tilde t_2}}$ at which electroweak
symmetry breaking is evaluated.\footnote{We assume that at the high scales the soft supersymmetry breaking
parameters are the same for all three generations; beyond that framework, e.g.~in models with inverted hierarchy of
soft supersymmetry breaking masses, two-loop contributions proportional to squark masses can drive $m_L^2$ to values
smaller than $m_E^2$, opening up a possibility for yet another example of sneutrino LOSP \cite{MBP}.}
%%%%%%%%%%%%%%%%
\begin{table}
\begin{center}
\begin{tabular}{|c|ccccc|ccccc|}
\hline
$m_S$ & $c_{E1}$ & $c_{L1}$ & $c_{L2}$  & $\tilde c_{EU}$ & $\tilde c_{LQ}$  & $c_{E1}$ & $c_{L1}$ & $c_{L2}$ &  $\tilde c_{EU}$ & $\tilde c_{LQ}$  \\
& \multicolumn{5}{|c|}{$Q=10^{14}\,\mathrm{GeV}$} & \multicolumn{5}{c|}{$Q=10^{16}\,\mathrm{GeV}$} \\
\hline
$500\,\mathrm{GeV}$ & 0.47 & 0.12 & 0.52 & $-0.0027$ & {$-0.0049$} &  0.62 & 0.15 & 0.64 & $-0.0038$ & {$-0.0060$} \\
$1000\,\mathrm{GeV}$ & 0.45 & 0.11 & 0.51 & $-0.0026$ & {$-0.0048$}  & 0.59 & 0.15 & 0.62 & $-0.0037$ &  {$-0.0059$}  \\
%lr & $c_{E,1}$ & $c_{L,1}$ & $c_{L,2}$  & $\tilde c_{E,U}$ & $\tilde c_{L,Q}$  & $c_{E,1}$ & $c_{L,1}$ & $c_{L,2}$ &  $\tilde c_{E,U}$ & $\tilde c_{L,Q}$  \\
%lr & \multicolumn{5}{|c|}{$Q=10^{14}\,\mathrm{GeV}$} & \multicolumn{5}{c}{$Q=10^{15}\,\mathrm{GeV}$} \\
%lr \hline
%lr $m_S=500\,\mathrm{GeV}$ & 0.47 & 0.12 & 0.52 & $-0.0027$ & {$-0.0049$} &  0.58 & 0.14 & 0.59 & $-0.0032$ & {$-0.0055$} \\
%lr $m_S=1000\,\mathrm{GeV}$ & 0.45 & 0.11 & 0.51 & $-0.0026$ & {$-0.0048$}  & 0.56 & 0.14 & 0.57 & $-0.0031$ &  {$-0.0054$}  \\
\hline
\end{tabular}
\end{center}
\caption{Numerical values of the coefficients $c_{E1}$, $c_{L1}$,
  $c_{L2}$, $\tilde c_{EU}$, $\tilde c_{LQ}$ in (\ref{eq:solrge})
  for two representative choices of the high scale $Q$ and of the EWSB
  mass scale $m_S=\sqrt{m_{\tilde t_1}m_{\tilde t_2}}$.  \label{t1} } 
\end{table}
%%%%%%%%%%%%%%%%
$D^2$ (denoted in literature also as $\mathcal{S}_0$) is defined as
%%%%%%%%%%%%%%%%
\begin{equation}
D^2 = \mathcal{S}_0=\mathrm{tr} \left[Y\mathbf{M}^2_{\mathrm{scalars},0}\right]=
m_{H_u}^2 - m_{H_d}^2+\mathrm{tr}\left[  \mathbf{m}_{Q,0}^2 -2
  \mathbf{m}_{U,0}^2 + \mathbf{m}_{D,0}^2 - \mathbf{m}_{L,0}^2
  +\mathbf{m}_{E,0}^2  \right] \,,  
\end{equation}
%%%%%%%%%%%%%%%%
where $\mathbf{m}_{S,0}^2$ are the $3\times3$ 
sfermion mass matrices at the high scale, $m_{H_u}^2$ and $m_{H_u}^2$
are the soft supersymmetry breaking masses of the Higgs doublets  at the high scale, 
 and $g_1\,(g_{1,0})$ is the
$U(1)_Y$ gauge coupling at the low (high) scale.  Leading corrections
arising due to the $\tau$ Yukawa couplings are denoted by
$\delta^2_{E,y_\tau}$ and $\delta^2_{L,y_\tau}$; for small and
moderate values of $\tan\beta$ they are small and their only role is
to make the third generation of sleptons slightly lighter than the
first two, but they can become important if the mass parameters
$\sqrt{m_{H_d}^2}$ at the high scale or the coefficient $A_\tau$ in
the trilinear coupling of staus are much larger than $\sqrt{m_L^2}$
and $\sqrt{m_E^2}$.  In the case when the colored particles are much
heavier than the sleptons, as is usually the case, one
should in principle include the leading two-loop contributions to the
RGEs in order to obtain $\mathcal{O}(10\,\mathrm{GeV})$ accuracy in mass
determination.

%Substituting (\ref{eq:solrge}) and (\ref{eq:solrgea}) into
%(\ref{eq:snucon}), we see that the sneutrino can be the LOSP in two
%(mutually not exclusive) cases.  Firstly, we can take $D^2<0$. In this
%case the sign differences in coefficients multiplying $D^2$ in
%(\ref{eq:solrge}) and (\ref{eq:solrgea}) can lead to
%$m_L^2<m_E^2$. This possibility is realized in the general NUHM and
%later we shall also discuss corresponding mass spectra in some detail.
%Note that more unified models such as CMSSM or NUHM1 (i.e.\ the NUHM
%with $m_{H_u}=m_{H_d}$) predict $D^2=0$.  
%%
%Then, at the high scale the
%parameter $m_L^2$ has to be the small relative to other soft
%supersymmetry breaking parameters.
%% or, alternatively, $m_E^2$ and $M_1^2$  must be both
%%sufficiently larger than $m_L^2$ and $M_2^2$.  
%%
%%In particular, 
%In
%models with universal gaugino masses, for which $M_2\approx 2M_1$, and
%a high scale $>10^{14}\,\mathrm{GeV}$, the sneutrino cannot be the
%LOSP, as it is always heavier than the bino. This leads us to the
%second possibility which is to relax gaugino mass universality. This
%possibility can naturally be realized in GGM models, leading to the
%sneutrino LOSP and we shall later present some representative examples
%mass spectra arising in such models.

Substituting (\ref{eq:solrge}) and (\ref{eq:solrgea}) into
(\ref{eq:snucon}), we see that the sneutrino can be the LOSP in two
(mutually not exclusive) cases. Note first that in models with $D^2
=0$ and universal gaugino masses, such as the CMSSM or the NUHM1 model
(i.e.\ the NUHM with $m_{H_u}=m_{H_d}$), for which $M_2\approx 2M_1$,
and a high scale $>10^{14}\,\mathrm{GeV}$, the sneutrino cannot be the
LOSP, since it is always heavier than the bino. We can then firstly
demand $D^2<0$, which gives $m_L^2 < M_1^2$, and then the sign
difference in the coefficients multiplying $D^2$ in (\ref{eq:solrge})
and (\ref{eq:solrgea}) can lead to $m_L^2<m_E^2$. This possibility is
realized in the general NUHM and later we shall also discuss the
corresponding mass spectra in some detail.  The second option is to
relax the gaugino mass universality. This possibility is naturally
realized in GGM models, leading to the sneutrino LOSP and we shall
later present some representative examples of mass spectra arising in
such models, as well.\footnote{Another way would be to assume large
  $m_{Q,0}$, since it would give a negative contribution to
  $m_L^2$. However, this would lead to large $\mu$, hence would
  increase the left-right mixing in the stau sector and would thus
  make the lighter stau lighter than the sneutrino.}

\subsection{Higgs boson mass of $126\,\mathrm{GeV}$}

Recent data from the LHC \cite{:2012gk,:2012gu} strongly suggest that the
lightest Higgs boson has a mass of approximately 126 GeV. 
As we mentioned in the Introduction, this implies a larger supersymmetry
breaking scale and implies non-trivial consequences for the
possibility of having sneutrino NLSP with gravitino LSP.

At one loop, the lightest Higgs boson mass can be approximated as \cite{Haber:1996fp}
%%%%%%%%%%%%%%%%
\begin{equation}
\label{eq:higgs}
m_h^2 \approx m_Z^2\cos 2\beta+\frac{3}{4\pi^2}\frac{m_t^4}{v^2} \left[ \log \frac{m_S^2}{m_t^2} + \frac{X_t^2}{m_S^2}\left(1-\frac{X_t^2}{12m_S^2}\right)\right] \, , 
\end{equation}
%%%%%%%%%%%%%%%%
where $v = 174\,\textrm{GeV}$, $m_S^2$ is the (defined above) product of the stop masses and
$X_t=A_t-\mu/\tan\beta$. It is well known (see e.g.\
\cite{Hall:2011aa,Heinemeyer:2011aa,Arbey:2011aa,Draper:2011aa,Carena:2011aa})
that consistency with the Higgs boson mass measurement at
$\sim126\,\mathrm{GeV}$ points toward large values of $m_S\simgt
\mathcal{O}(1)\,\mathrm{TeV}$ and values of $X_t$ maximizing the
second term in the square bracket in (\ref{eq:higgs}), with largest values achieved for \ $X_t\sim \pm\sqrt{6}m_S$. (The
other option of increasing $m_S$ so that the logarithmic correction in
(\ref{eq:higgs}) gives  the whole necessary contribution is less natural
as it requires very heavy stops.) A solution of the one-loop
MSSM RGEs gives \cite{Carena:1996km}:
%\begin{eqnarray}
%\label{eq:solat}
%A_t &\approx& (1-y)A_{t,0}-(\xi_u-y\hat\xi)m_{1/2} \\
%\label{eq:solmu}
%\mu^2 & \approx & \frac{1}{2}y(1-y)(A_{t,0}^2-2\hat\xi A_{t,0}m_{1/2})+\left[ -\eta_{H_u}+ y(\hat\eta-\frac{1}{2}y\xi^2)\right]m_{1/2}^2+\textrm{other terms} \\
%\label{eq:solq}
%m_Q^2 & \approx & -\frac{1}{6}y(1-y)(A_{t,0}^2-2\hat\xi A_{t,0}m_{1/2})+\left[ \eta_{Q}-\frac{1}{3} y(\hat\eta-\frac{1}{2}y\xi^2)\right]m_{1/2}^2+\textrm{other terms}\\
%\label{eq:solu}
%m_U^2 & \approx & -\frac{1}{3}y(1-y)(A_{t,0}^2-2\hat\xi A_{t,0}m_{1/2})+\left[ \eta_{U}- \frac{2}{3}y(\hat\eta-\frac{1}{2}y\xi^2)\right]m_{1/2}^2+\textrm{other terms} \, . 
%\end{eqnarray}
%OR
%%%%%%%%%%%%%%%%
\begin{eqnarray}
\label{eq:solat}
A_t &=& c^{A_t}_{A} A_{t,0}-c^{A_t}_{1/2}m_{1/2} \\
\label{eq:solmu}
\mu^2 & \approx & 3c^{\mu}_A A_{t,0}^2-3c^\mu_{A,1/2} A_{t,0}m_{1/2}+c^\mu_{1/2}m_{1/2}^2+\ldots \\
\label{eq:solq}
m_Q^2 & \approx & -c^\mu_AA_{t,0}^2+c^\mu_{A,1/2} A_{t,0}m_{1/2}+c^Q_{1/2}m_{1/2}^2+\ldots \\
\label{eq:solu}
m_U^2 & \approx & -2c^\mu_AA_{t,0}^2+2c^\mu_{A,1/2} A_{t,0}m_{1/2}+c^U_{1/2}m_{1/2}^2+\ldots \, . 
\end{eqnarray}
%%%%%%%%%%%%%%%%
%%%%%%%%%%%%%%%%
 \begin{table}
\begin{center}
\begin{tabular}{|c|ccccccc|}
\hline
coefficient & $c^{A_t}_A$ & $c^{A_t}_{1/2}$ & $c^\mu_A$ & $c^\mu_{A,1/2}$ & $c^\mu_{1/2}$ & $c^Q_{1/2}$ & $c^U_{1/2}$ \\
\hline
range & $\sim 0.4$ & $\sim 2$ & $\sim 0.04$ & $\sim 0.1$  & $\sim 3$ & $\sim 3$ & $1-2$ \\
\hline
\end{tabular}
\end{center}
\caption{Approximate values of the coefficients $c^\alpha_\beta$ in (\ref{eq:solat})-(\ref{eq:solu}) for $m_S$ varying from
1 to 5 TeV and two patterns of gaugino masses at the high scale $Q=2\times10^{16}\,\mathrm{GeV}$: universal gaugino case and $M_{1,0},M_{2,0}\ll M_{3,0}\equiv m_{1/2}$. \label{t2}}
\end{table} 
%%%%%%%%%%%%%%%%
At one loop the values of the numerical coefficients $c^\alpha_\beta$
can be expressed as functions of the gauge and top Yukawa couplings.
In Table \ref{t2} we indicate typical values of these coefficients 
for different choices of $M_S$ and gaugino mass patterns.

For brevity, in (\ref{eq:solmu})-(\ref{eq:solu}) only the terms depending
on the high-scale parameters $m_{1/2}$ and $A_{t,0}$ are shown, as
they suffice for the following argument.  From
(\ref{eq:solat})-(\ref{eq:solmu}) it is obvious that the easiest
way of obtaining a large negative $X_t$ is to make the gluino rather
heavy; increasing $A_{t,0}$ by an equal amount is about five times
less effective and may threaten to make the stops tachyonic. However,
it should be kept in mind that a large $m_{1/2}$ tends to make $|\mu|$
large; it is of no particular consequence for $X_t$, as $\mu$ enters
this quantity multiplied by $1/\tan\beta$, but a large $|\mu|$
additionally increases left-right sfermion mixing, which, as we
discussed in Section \ref{sec:snulow} tends to make charged sleptons
lighter than sneutrinos (for fixed sfermion masses at the high scale).

If the soft supersymmetry breaking mass parameters $A_t$, $m_Q^2$ and
$m_U^2$ are dominated by the RGE contributions from the gluino mass,
one obtains $X_t/m_S \approx X_t/\sqrt{m_Qm_U}\sim -1$, which is not
very close to the maximal stop mixing scenario \cite{Dermisek:2006ey},
optimal for a large Higgs boson mass (the second term in the square
bracket in (\ref{eq:higgs}) is $\sim2$ times smaller than its maximal
value).  A Higgs boson mass of 126 GeV can be then obtained
either by assuming a rather large $m_{1/2}$, or by taking a large
negative $A_{t,0}$, preferably $A_{t,0}\sim-(1-3)m_{1/2}$
\cite{Brummer:2012ns},  or else by admitting tachyonic stops at high scales
\cite{Draper:2011aa}, which we shall not pursue further here.

We are therefore led to the conclusion that a Higgs boson mass of 126
GeV puts an important constraint on the possibility of sneutrino LOSP 
by implying a higher scale of supersymmetry breaking. 
Lower bounds from direct SUSY searches are consistent with this trend
but currently not yet as strong.  A large Higgs boson
mass favors large negative $A_t$ which is usually correlated with
large $A_\tau$ via RG running; this increases left-right mixing in the
charged slepton sector and makes a stau lighter than the
sneutrino. Such a large $A_t$ most easily originates from a large
$m_{1/2}$ (or a combination of slightly smaller $m_{1/2}$ and a large
negative $A_{t,0}$), which increases $\mu$, thereby also increasing
left-right mixing in the charged slepton sector. In the following we
will illustrate within two SUSY models employing different
supersymmetry breaking mechanisms, and both allowing sneutrino LOSP, how a
large value of $m_{1/2}$ implied by a heavy Higgs boson leads to strong
constraints on the reheating temperature resulting from BBN bounds.

%%%%%%%%%%%%%%%%
\subsection{Bound on the reheating temperature
  from BBN}
%%%%%%%%%%%%%%%%

For gravitinos with a mass significantly smaller than the Fermi scale,
their present abundance resulting from scatterings in thermal plasma
\cite{Bolz:2000fu,Pradler:2006qh,Rychkov:2007uq} can be approximated by \cite{Olechowski:2009bd}:
%%%%%%%%%%%%%%%%
\beq
\Omega_{\tilde G}^\mathrm{TP} h^2 \approx \left(
  \frac{T_\mathrm{R}}{10^8\,\mathrm{GeV}}\right)\left(
  \frac{1\,\mathrm{GeV}}{m_{\tilde G}} \right) \sum_{r=1}^3 \gamma_r
\left( \frac{M_r}{900\,\mathrm{GeV}} \right)^2 \, , 
\eeq
%%%%%%%%%%%%%%%%
where $m_{\tilde G}$ is the gravitino mass, $M_r$ denote gaugino mass
parameters at the low scale and the coefficients $\gamma_r$ can be
calculated from 1-loop RGEs for the gaugino masses and gauge couplings: they can be
evaluated for $T_\mathrm{R}=10^9\,(10^7)\,\mathrm{GeV}$ as
$\gamma_3=0.50\,(0.67)$, $\gamma_2=0.51\,(0.49)$,
$\gamma_1=0.20\,(0.15)$, for the gluino masses of 900 GeV.  It is easy
to read from this estimate that with $m_{1/2}\sim 1\,\mathrm{TeV}$ and
$m_{\tilde G}=100\,\mathrm{GeV}$ the observed dark matter abundance
implies a reheating temperature of $T_R\sim 5\times
10^8\,\mathrm{GeV}$, which is close to minimal values $\sim2\times
10^9\,\mathrm{GeV}$ ($\sim2\times 10^8\,\mathrm{GeV}$) required by
simple models of thermal leptogenesis with zero (thermal) initial
abundance of the lightest right-handed neutrinos and sneutrinos
\cite{Giudice:2003jh,Antusch:2006gy}.

The lifetime of sneutrino LOSP can be approximated as
%%%%%%%%%%%%%%%%
\beq
\label{tau}
\tau_\mathrm{NLSP} = \left(5.9\times 10^4\, {\rm sec}\right)
\left(\frac{m_{\tilde G}}{1\,\mathrm{GeV}}\right)^2 \left(
  \frac{100\,\mathrm{GeV}}{m_\mathrm{NLSP}}\right)^5 \left(
  1-\frac{m_{\tilde G}^2}{m_\mathrm{NLSP}^2}\right)^{-4} \, , 
\eeq
%%%%%%%%%%%%%%%%
which can easily be of the order of $10^5-10^7\,\mathrm{sec}$. 
%lr For low $m_{\tilde G}$ and sneutrino mass of $250-500$ GeVis of the order of $10^5-10^7\,\mathrm{sec}$. 
For such long lifetimes it is then possible that hadro-dissociation
processes induced by a subdominant decay process of sneutrino LOSP
where a quark-antiquark pair is produced can alter the BBN predictions
beyond the current observational uncertainties. We shall study this
issue in the following Section, by numerically analyzing
representative examples in two models of supersymmetry breaking which
allow for sneutrino LOSP.

\section{Numerical analysis}

As we have argued in Section \ref{sec:snurge}, models of supersymmetry
breaking at the high scale allow a sneutrino LOSP only if at least one
of the two conditions: 
%$\mathrm{tr}(Y\mathbf{M}_\mathrm{scalars}^2)=0$ 
$D^2=0$
or
$M_1:M_2:M_3=\alpha_1:\alpha_2:\alpha_3$ is violated at the high scale.
(At 1 loop these conditions are renormalization group invariants,
hence they can be evaluated at any scale.)  A violation of the former
is manifest in the NUHM while  the latter condition can be
satisfied in many ways. A recently considered scenario is the GGM where
it is assumed that the hidden and the messenger sectors in models of gauge
mediation can be more complicated than what is required in a minimal  theory. 
(The soft supersymmetry breaking parameters in both models are given in the Appendix.) 
%In this Section, we shall
%numerically explore the regions of the corresponding parameter spaces
%where the sneutrino can be the LOSP.

%%%%%%%%%%%%%%%%
% table was here
%%%%%%%%%%%%%%%%

%%%%%%%%%%%%%%%%
\subsection{The NUHM}
%%%%%%%%%%%%%%%%

%%%%%%%%%%%%%%%%
\begin{table}
\begin{center}
\begin{tabular}{|c|cc|cc|}
\multicolumn{5}{c}{All scans: $m_{H_u}=500\,\mathrm{GeV}$, $m_{H_d}=4000\,\mathrm{GeV}$, $\mu>0$} \\
\hline
Case & \multicolumn{2}{|c|}{varied parameters} & \multicolumn{2}{|c|}{fixed parameters} \\
\hline
1 & $m_0$ & $m_{1/2}$ & $A_0=-3000\, \mathrm{GeV}$ & $\tan\beta=10$ \\
2 & $A_0$ & $m_{1/2}$ & $m_0=300\, \mathrm{GeV}$ & $\tan\beta=10$ \\
3 & $A_0$ & $\tan\beta$ &  $m_0=300\, \mathrm{GeV}$ & $m_{1/2}=1200\,\mathrm{GeV}$ \\
4 & $m_{1/2}$ & $\tan\beta$ & $m_0=300\, \mathrm{GeV}$ & $A_0=-3000\, \mathrm{GeV}$ \\
\hline
\end{tabular}
\end{center}
\caption{Description of scans over the parameters in the NUHM
  presented in Figures \ref{fn} and \ref{fnn}. \label{tsn}} 
\end{table}
%%%%%%%%%%%%%%%%

Armed with the above analytical considerations, we will now identify
regions of the NUHM parameter space where the sneutrino is the
LOSP. We will determine if these solutions are consistent with the
Higgs mass, low-energy observables and early Universe. The details of
the scans are given in Table \ref{tsn}. In our numerical work we used
{\tt suspect} \cite{Djouadi:2002ze} to solve the renormalization group
equations and calculate mass spectra, {\tt micrOMEGAs}
\cite{Belanger:2006is} for the LOSP relic abundance and {\tt SuperIso}
\cite{Mahmoudi:2008tp} for flavor observables.

%%%%%%%%%%%%%%%%

\begin{figure}[t]
\begin{center}
\includegraphics*[height=7cm, width=7cm, trim= 0mm 0mm 0mm 0mm]{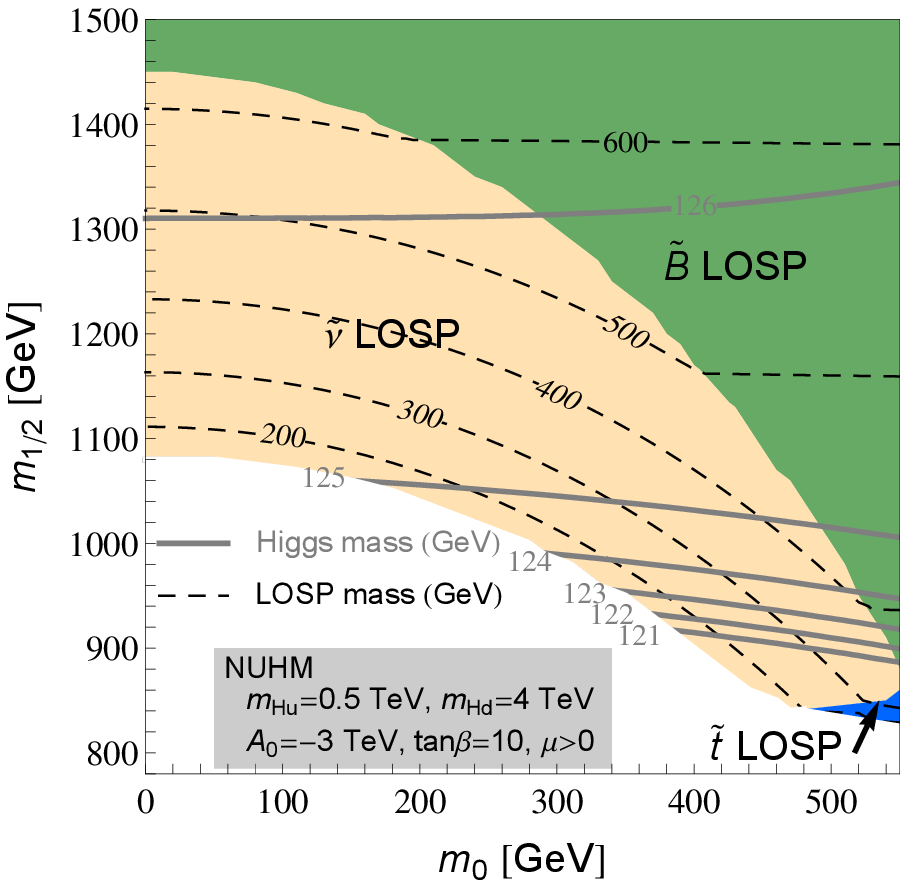}
\hspace{0.5cm}
\includegraphics*[height=7cm, width=7cm, trim= 0mm 0mm 0mm 0mm]{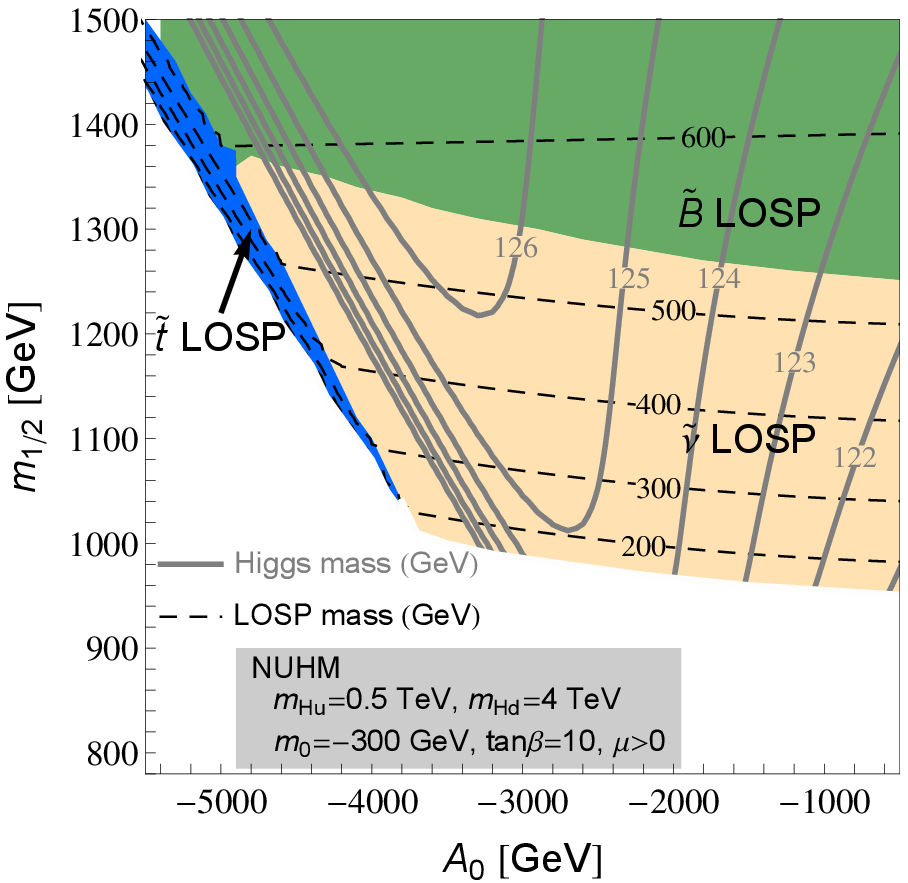}
\end{center}
\caption{
Slices of the NUHM parameter space: $m_0$ vs $m_{1/2}$ (left panel)
and $A_0$ vs $m_{1/2}$ (right panel) with the values 
of $m_{H_u}=500\,\mathrm{GeV}$, $m_{H_d}=4000\,\mathrm{GeV}$ fixed at
the unification scale and $\tan\beta=10$, $\mu>0$. Contours of
constant LOSP (Higgs boson) masses are shown as dashed (solid)
lines. Unphysical regions are marked in white. 
\label{fn}}
\end{figure}

\begin{figure}
\begin{center}
\includegraphics*[height=7cm, width=7cm, trim= 0mm 0mm 0mm 0mm]{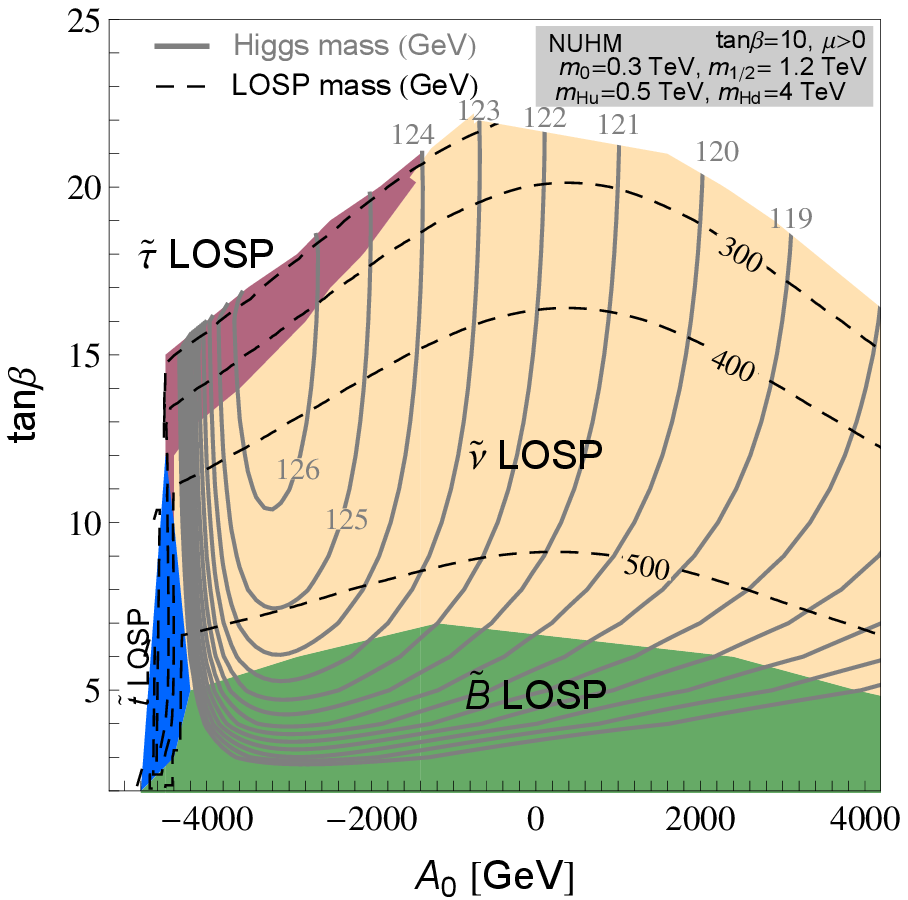}
\hspace{0.5cm}
\includegraphics*[height=6.99cm, width=7cm, trim= 0mm 0mm 0mm 0mm]{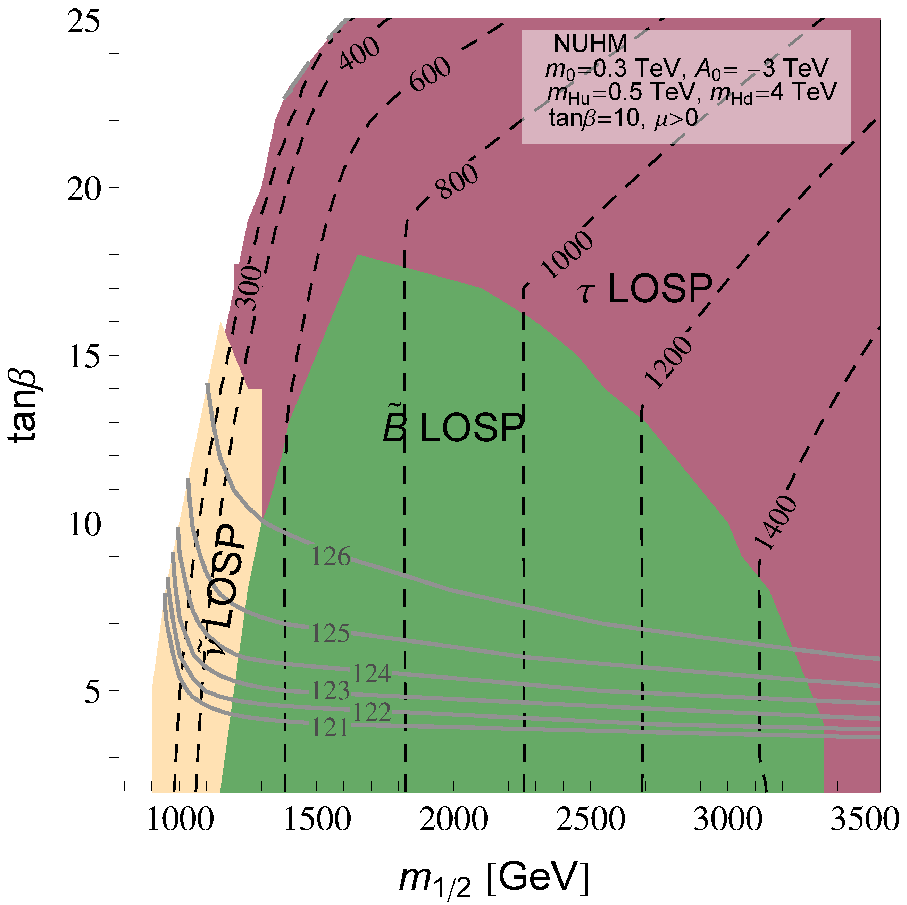}
\end{center}
\caption{
Slices of the NUHM parameter space: $A_0$ vs $\tan\beta$ (left panel) and $m_{1/2}$ vs $\tan\beta$ (right panel) with the values
of $m_0=300\,\mathrm{GeV}$, $m_{H_u}=500\,\mathrm{GeV}$,
$m_{H_d}=4000\,\mathrm{GeV}$ fixed at the unification scale and
$\mu>0$. Contours of constant LOSP (Higgs boson) masses are shown as
dashed (solid) lines. Unphysical regions are marked in white.  
 \label{fnn}}
\end{figure}

In Figures \ref{fn} and \ref{fnn} we present the LOSP identity and its
mass, as well as the mass of the Higgs boson.
In both panels of Figure \ref{fn} and in the right panel of Figure
\ref{fnn} the sneutrino LOSP region is bounded from above at large enough
values of $m_{1/2}$. This can be easily understood since $m_L^2$ is a
much faster growing function of $m_{1/2}$ than $M_1^2$ which is the bino
mass squared. At fixed $m_{1/2}$  and increasing $m_0$,
sfermion masses grow and they eventually become
larger than the bino mass, which explains the bending of the boundary
between the sneutrino and bino LOSP regions in the left panel of
Figure \ref{fn}. A negative contribution to $m_L^2$, which is  proportional to
$D^2$, has to be overcome by some other positive contributions proportional
to $m_0$ or $m_{1/2}$; otherwise we find unphysical
regions (marked white) with tachyonic sleptons.  It should also be
mentioned that the negative contribution to $m_U^2$ (proportional to
$D^2$) is larger by a factor of 3/2 than that to $m_L^2$; the former
parameter also receives a much larger renormalization group correction
proportional to $m_{1/2}^2$ than the latter. As a consequence, for
sufficiently small values of $m_{1/2}$ the lighter stop becomes lighter than
the sleptons; the corresponding
region of stop LOSP is visible in the left panel of Figure \ref{fn}.  All these
effects leads to a lower bound on $m_{1/2}$; in our scans we find no
sneutrino LOSP models for $m_{1/2}<800\,\mathrm{GeV}$, which,
as we shall discuss later, has the important consequences for
the maximum reheating temperature.  As it can be
seen in the right panels of Figures \ref{fn} and \ref{fnn}, for
$\mu>0$ large negative values of $A_0$ result in large off-diagonal
entries in the stop mass matrix and lead to very light and even
tachyonic stops.  The appearance of the bino LOSP region in Figure
\ref{fnn} results from the $\tau$-Yukawa effect in the renormalization
group equations: in the leading logarithm approximation, the
quantities $\delta^2_{E,y_{tau}}$ in (\ref{eq:solrge}) and
$\delta^2_{L,y_{\tau}}$ in (\ref{eq:solrgea}) can be approximated by

%%%%%%%%%%%%%%%%%
%%%%%%%%%%%%%%%%

%%%%%%%%%%%%%%%%
%%%%%%%%%%%%%%%%
%%%%%%%%%%%%%%%%
\begin{equation}
\delta^2_{E,y_{\tau}} \approx 2\delta^2_{l,y_{tau}} \approx -\frac{1}{4\pi^2}y_\tau^2(m_{H_d}^2+A_0^2) \log\left(\frac{M_\mathrm{GUT}}{m_S}\right) \,,
\end{equation}
%%%%%%%%%%%%%%%%
where $M_\mathrm{GUT}$ is the unification scale at which $m_{H_d}$ and
$y_\tau$ are evaluated here. As $y_\tau$ increases with growing
$\tan\beta$, we see that moderate values of $\tan\beta$ actually help
one of the sleptons to become the LOSP.

Unsurprisingly, for fixed $A_0=-3000\,\mathrm{GeV}$, we find a lower
bound $m_{1/2}\simgt 1\,\mathrm{TeV}$ resulting from the requirement
that the Higgs boson mass exceeds 122 GeV, the value that we adopt as
a conservative lower bound on the observable. For fixed $m_0$, the shape
of the constant Higgs boson mass contours in the $(A_0,m_{1/2})$ plane
agrees with the requirements for maximal stop mixing
\cite{Brummer:2012ns}.
For all the points shown in Figures \ref{fn} and \ref{fnn} the
low-energy observables lie within a conservative 95\% CL range quoted
in Table \ref{tb}. 
As in many unified models, 
the supersymmetric contribution to
$\delta a_\mu$ is too low
to explain the observed anomaly \cite{Benayoun:2012wc}.
We also checked that for all the points of interest squark masses of the first and second generations are well
above 1400 GeV, required by the LHC data \cite{pdg_review}. In the stop LOSP regions,
stop masses are often much smaller that $\sim 450\,\mathrm{GeV}$ which
is the lower limit from the LHC, but these regions are disfavored
anyway, because the Higgs boson mass drops below 120 GeV there.

%%%%%%%%%%%%%%%%
\begin{table}
\begin{center}
\begin{tabular}{|c|cccc|}
\hline
observable & 
$\mathrm{BR}(b\to s\gamma)$ \cite{slac} &
$\mathrm{BR}(B_u\to \tau\nu_\tau)$ \cite{HFAV} &
$\mathrm{BR}(B_s\to \mu^+\mu^-)$ \cite{LHCb} &
%$\delta a_\mu$ \cite{pdg} \\
$\Delta M_{B_s}$ \cite{pdg2}\\
\hline
lower bound &
$2.8\times 10^{-4}$ &
$0.7\times 10^{-4}$ &
$0.7\times 10^{-9}$ &
$12.9\,\mathrm{ps}^{-1}$
\\
upper bound &
$4\times 10^{-4}$ &
$2.7\times 10^{-4}$ &
$6.3\times 10^{-9}$ &
$22.5\,\mathrm{ps}^{-1}$ \\
\hline
\end{tabular}
\end{center}
\caption{95\% CL bounds for selected low-energy flavor and
  electromagnetic observables. Both experimental and theoretical
  errors have been taken into account. 
 \label{tb}}
\end{table}
%%%%%%%%%%%%%%%%
 
%%%%%%%%%%%%%%%%
 \begin{table}
\begin{center}
\begin{tabular}{|c|cccc|}
\hline
observable &
D/H & ${}^3\mathrm{He}/\mathrm{D}$ & $Y_\mathrm{p}$ & ${}^6\mathrm{Li}/^7\mathrm{Li}$ \\
\hline
lower bound
&
$1.2\times 10^{-5}$ &
&
not applied &
\\
upper bounds
&
$4\times 10^{-5}$/$5.3\times 10^{-5}$ &
$1.5$ &
0.26 &
0.1/0.66\\
 & & & & (stringent/conservative)\\
\hline
\end{tabular}
\end{center}
\caption{95\% CL BBN bounds based on \cite{Jedamzik:2006xz}. 
The observables are ratios of the element abundances,
with the obvious exception of $Y_\mathrm{p}$ which is ${}^4\mathrm{He}$ mass fraction.
The lower limit on $Y_p$ is irrelevant for constraining the abundance of decaying particles. The upper
limit on D/H represents a compromise between the commonly used average 
of the best determinations of this quantity and
the large spread of the individual results. The use of two bounds for ${}^6\mathrm{Li}/^7\mathrm{Li}$, a stringent
and a conservative one, reflects the uncertainty in estimating the efficiency of production/destruction of this element
in stellar environment.
 \label{tbb}}
\end{table}
%%%%%%%%%%%%%%%%

\begin{figure}
\begin{center}
\includegraphics*[height=6.8cm, width=7cm, trim= 0mm 0mm 0mm 0mm]{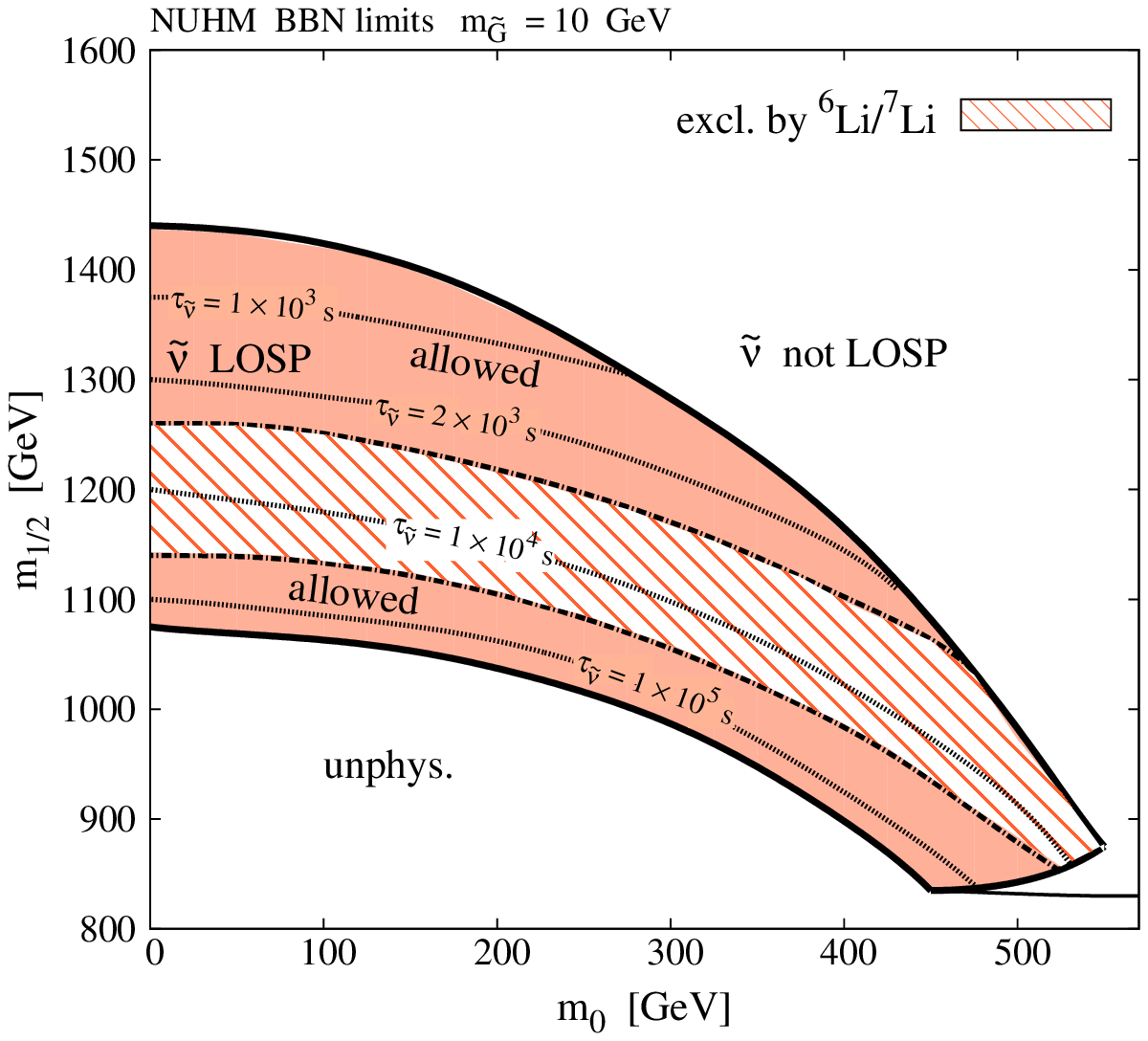}
\hspace{0.5cm}
\includegraphics*[height=6.8cm, width=7cm, trim= 0mm 0mm 0mm 0mm]{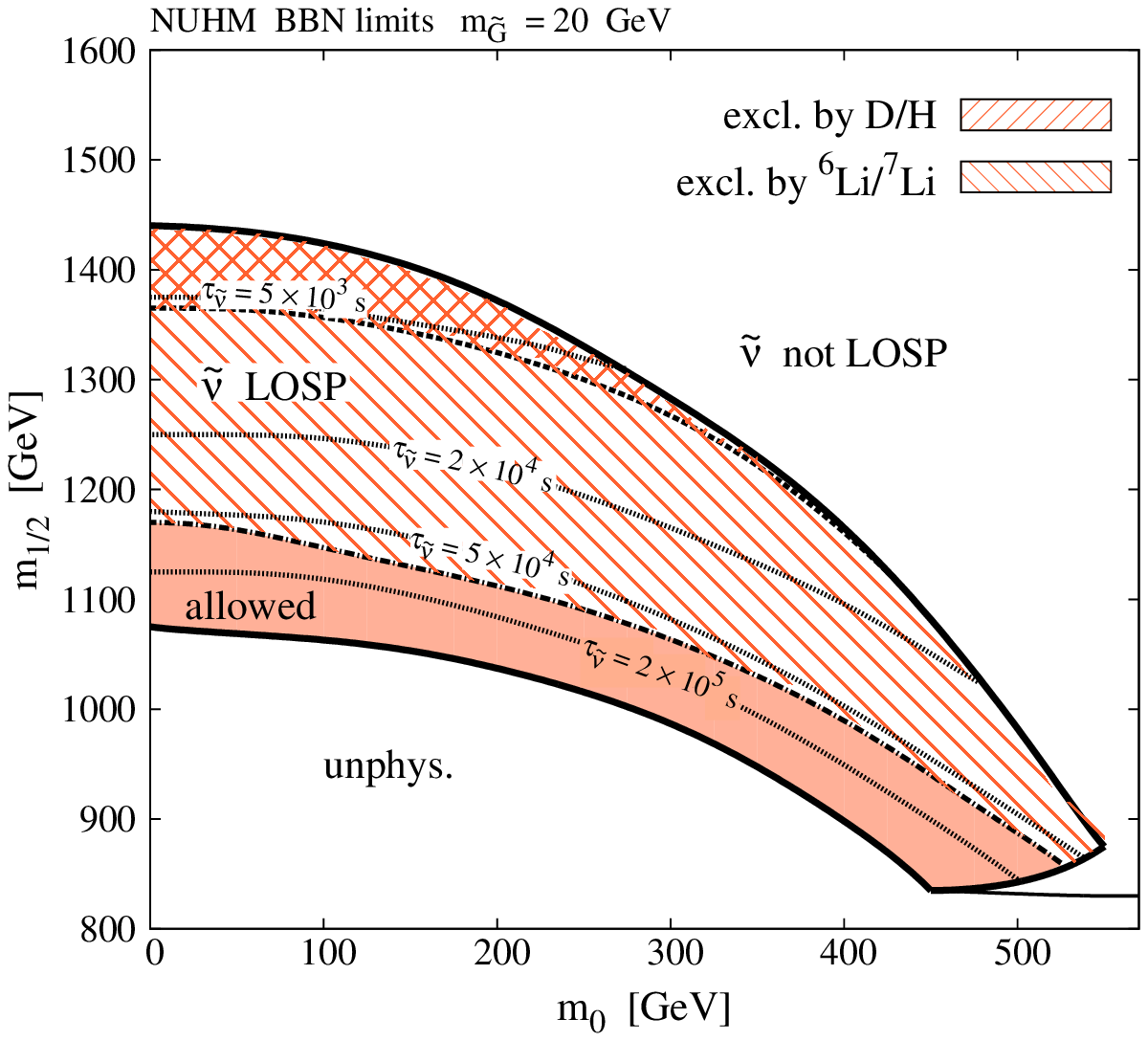}
\end{center}
\caption{
BBN bounds for the sneutrino LOSP region in the NUHM shown in the left panel of Figure \ref{fn}
for the values of gravitino mass of 
$m_{\tilde G}=10\textrm{ and }20\,\mathrm{GeV}$. For ${}^6\mathrm{Li}/^7\mathrm{Li}$
the stringent limit was used; the conservative limit does not constrain the parameter space.
 \label{fb1}}
\end{figure}

\begin{figure}
\begin{center}
\includegraphics*[height=6.8cm, width=7cm, trim= 0mm 0mm 0mm 0mm]{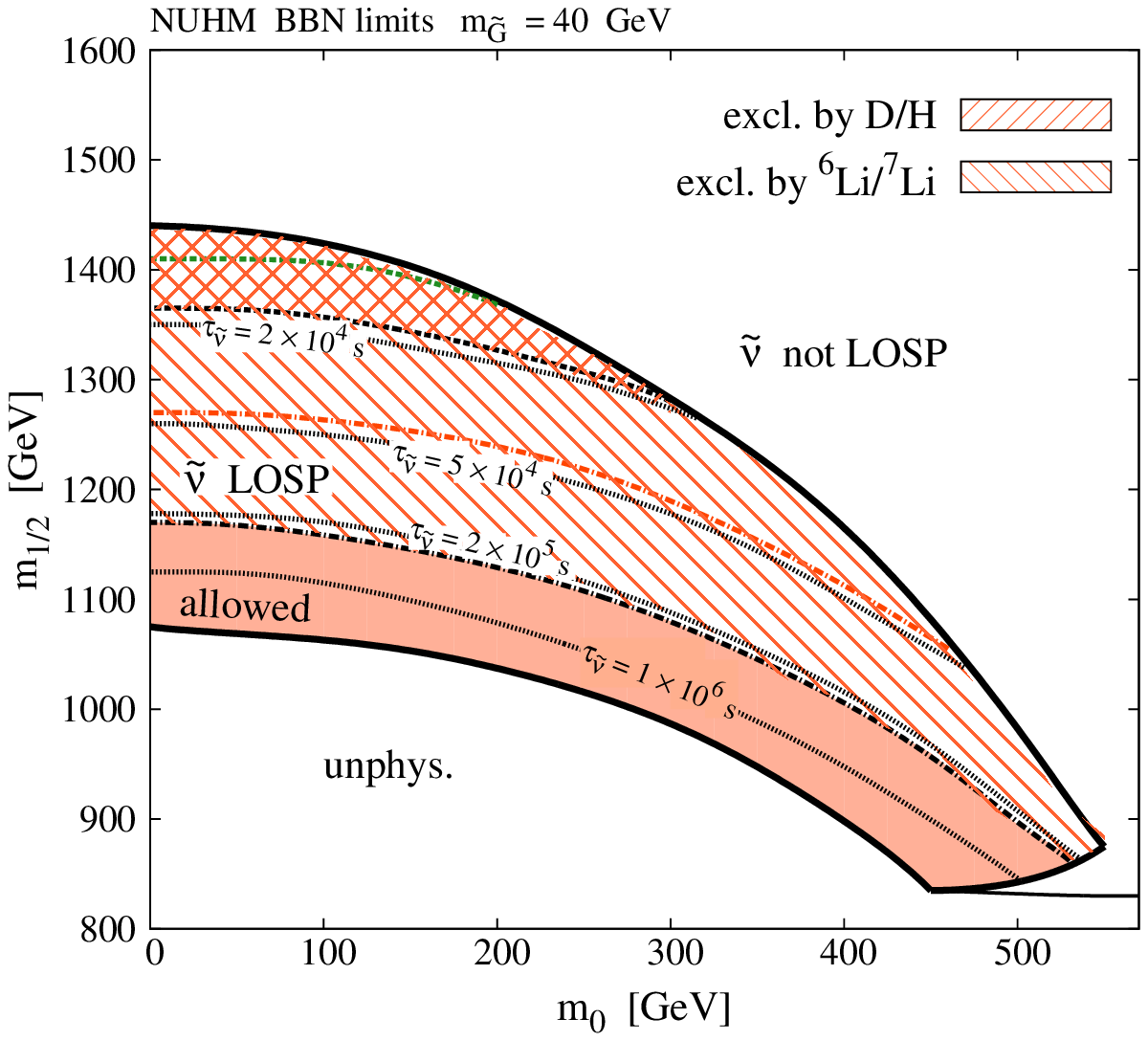}
\hspace{0.5cm}
\includegraphics*[height=6.8cm, width=7cm]{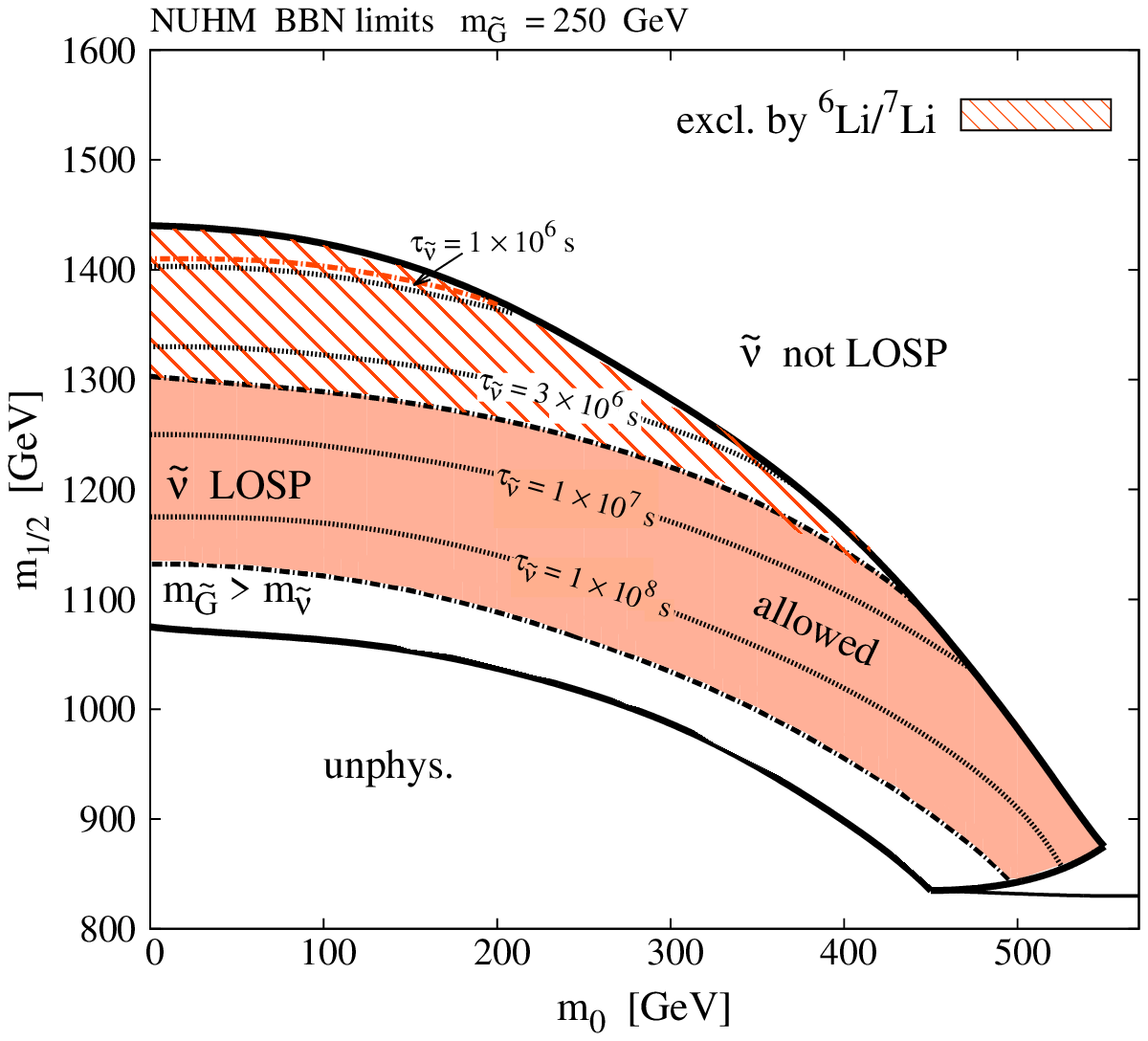}
\end{center}
\caption{
BBN bounds for the sneutrino LOSP region in the NUHM  shown in the left panel of Figure \ref{fn}
for the values of gravitino mass of $m_{\tilde G}=40\textrm{ and }250\,\mathrm{GeV}$.
For ${}^6\mathrm{Li}/^7\mathrm{Li}$
the stringent limit was used;
 the boundary of the excluded region with the more
conservative constraint for ${}^6\mathrm{Li}/{}^7\mathrm{Li}$ is represented by a red dash-dotted line. 
The dotted green line in the left panel shows the change in the lower boundary of the region excluded
by D/H if a more conservative limit $5.3\times 10^{-5}$ is used \cite{Jedamzik:2006xz}.
 \label{fb11}}
\end{figure}

For the region  with sneutrino LOSP shown in the
left panel of Figure \ref{fn} we calculate the abundances of light
elements following the method outlined in \cite{Jedamzik:2006xz} and
apply the observational limits shown in Table \ref{tbb}.  A
representative sample of our results is shown in Figures \ref{fb1} and
\ref{fb11}.  We find no constraints for the gravitino masses smaller
than 7.5 GeV. At $m_{\tilde G}=10\,\mathrm{GeV}$ a part of the
parameter space corresponding to
$m_{\tilde\nu}\simgt500\,\mathrm{GeV}$ or, equivalently, to
$\tau_{\tilde\nu}\simgt 10^3\,\mathrm{s}$, is excluded because of too
large D/H abundance. For all values of $m_{\tilde G}$ the bounds from
${}^6\mathrm{Li}/{}^7\mathrm{Li}$ are always more stringent than the
D/H bounds. A further increase of $m_{\tilde G}$ does not change this
picture much, until the gravitino becomes degenerate with the
sneutrino, which introduces a strong phase-space enhancement of the sneutrino lifetime. 
This is illustrated in the case with $m_{\tilde
  G}=250\,\mathrm{GeV}$, for which the BBN bounds, while still
dominated by ${}^6\mathrm{Li}/{}^7\mathrm{Li}$, become weaker.

\begin{figure}
\begin{center}
\includegraphics*[height=7cm, width=7cm, trim= 0mm 1mm 0mm 0mm]{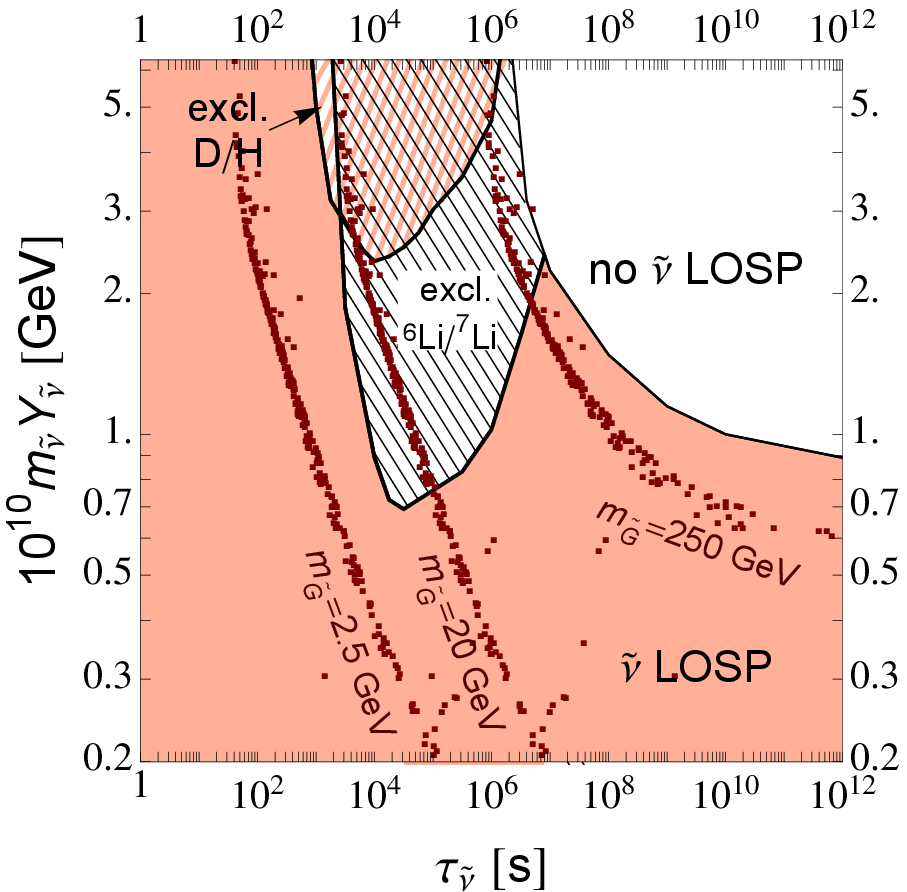}
\hspace{0.5cm}
\includegraphics*[height=7cm, width=7cm, trim= 0mm 0mm 0mm 2mm]{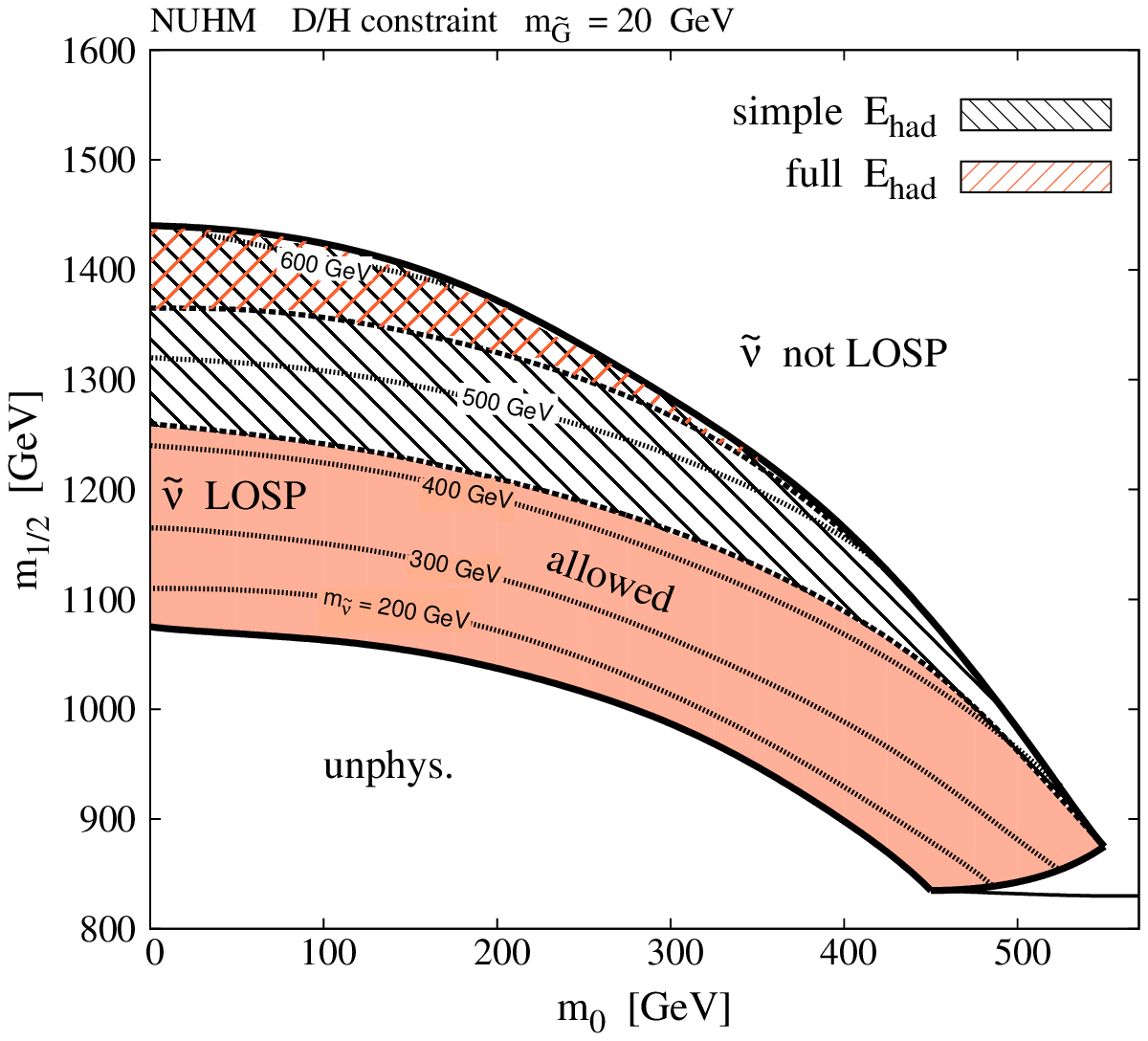}
\end{center}
\caption{
Left panel: BBN constraints shown in the $\tau_{\tilde\nu}$ vs $m_{\tilde\nu}Y_{\tilde\nu}$ plane for the sneutrino LOSP region
shown in the left panel of Figure \ref{fn}. Dots show the results of our scan with fixed $m_{\tilde G}=2.5$, 20 and 250 GeV.
Right panel: The impact of different estimates of hadronic energy release on the D/H bounds for $m_{\tilde G}=20\,\mathrm{GeV}$. For the excluded region marked `simple Ehad' an approximation $E_\mathrm{had}=(m_{\tilde\nu}-m_{\tilde G})/3$ was used, while the excluded region marked `full Ehad' corresponds to a computation of $E_\mathrm{had}$ involving integration over the full 4-body phase space. 
 \label{fb3a}}
\end{figure}

A closer look at the actual predictions for D/H in the considered
parameter range of the NUHM  reveals that even at points
consistent with the allowed bounds, the abundance of D is
altered with respect to the standard BBN value. 
It is also quite
sensitive to the hadronic energy release: if we approximated it as
$(m_{\tilde\nu}-m_{\tilde G})/3$, as is often done in the literature,
instead of calculating the energy of the $q\bar q$ pair produced in
the sneutrino decay, then with the conservative
${}^6\mathrm{Li}/{}^7\mathrm{Li}$ limit the lower boundary of the
respective excluded region in the right panel of Figure \ref{fb1}
would shift downwards by as much as $\sim100\,\mathrm{GeV}$ (see
the right panel of Figure \ref{fb3a}). In
  other words, one would significantly underestimate the sneutrino
  LOSP region allowed by the constraint.

In order to understand better the origin of the BBN constraints, we
first project all the analyzed points onto the $\tau_{\tilde\nu}$ vs
$m_{\tilde\nu}Y_{\tilde\nu}$ plane; this is shown in the left panel of Figure
\ref{fb3a}.  We also show there the bounds from the abundances of
those light elements that constrain parameter space regions with
sneutrino LOSP.  Since
$\Omega_{\tilde\nu}h^2$ is roughly proportional to $m_{\tilde\nu}^2$
(neglecting the opening of additional annihilation channels for
increased $\tilde\nu$ masses) and since for $m_{\tilde G}\ll
m_{\tilde\nu}$ the sneutrino lifetime scales as $\tau_{\tilde\nu}\propto m_{\tilde
  G}^2m_{\tilde\nu}^{-5}$, it is easy to understand why, with
increasing $m_{\tilde G}$, the constraints from D/H and
${}^6\mathrm{Li}/{}^7\mathrm{Li}$ first appear, next tighten up and
then eventually become weaker. As can be also easily seen from (\ref{tau}), a
partial degeneracy between $m_{\tilde G}$ and $m_{\tilde\nu}$ causes
a much larger increase of $\tau_{\tilde\nu}$  than the simple power law
above implies, hence the BBN bounds become correspondingly weaker.
Those features can easily be seen for the results of our scan with a
three values of fixed gravitino mass of $m_{\tilde G}=2.5,\,20,\,250\,\mathrm{GeV}$,
as
a band of dark red dots. 
%At large $m_{\tilde \nu}$ this band becomes
%steeper, which is related to a partial mass degeneracy between the
%sneutrino and the bino, and the resulting increase in the LOSP relic
%abundance due to coannihilations 
%(the average annihilation cross-section does not change, but the
%number of states to annihilate grows; quite a difference with the well-known bino-stau coannihilation
%scenario in the MSSM). 

One may worry that for long sneutrino lifetimes, $\tau_{\tilde\nu}>10^7\,\mathrm{s}$, the electromagnetic showers
produced in scatterings of energetic neutrinos from sneutrino decays off neutrinos of cosmic background can affect
the BBN by altering the ${}^3\mathrm{He/H}$ abundance \cite{Kawasaki:1994bs,Kanzaki:2007pd}. In order to verify this we determined 
that the exclusion plots from \cite{Kawasaki:1994bs} do not provide additional constraints on our parameter space; we also interpolated the exclusion plots from \cite{Kanzaki:2007pd} in $(m_{\tilde\nu},\tau_{\tilde\nu},B_h)$ plane and found
 no significant impact from ${}^3\mathrm{He/D}$. 

\begin{figure}
\begin{center}
\includegraphics*[height=6.8cm, width=7cm, trim= 0mm 0mm 0mm 0mm]{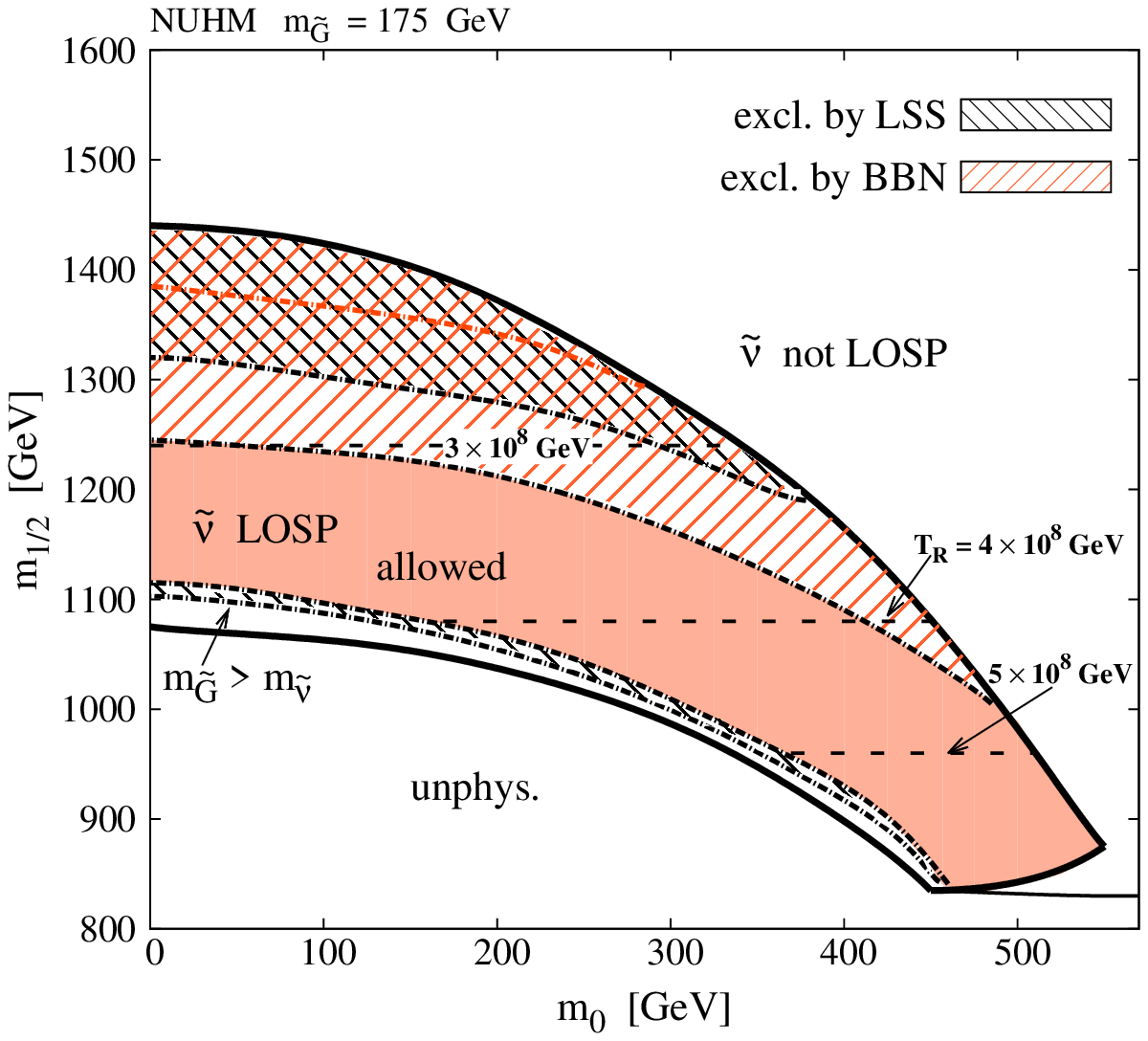}
\hspace{0.5cm}
\includegraphics*[height=6.8cm, width=7cm, trim= 0mm 0mm 0mm 0mm]{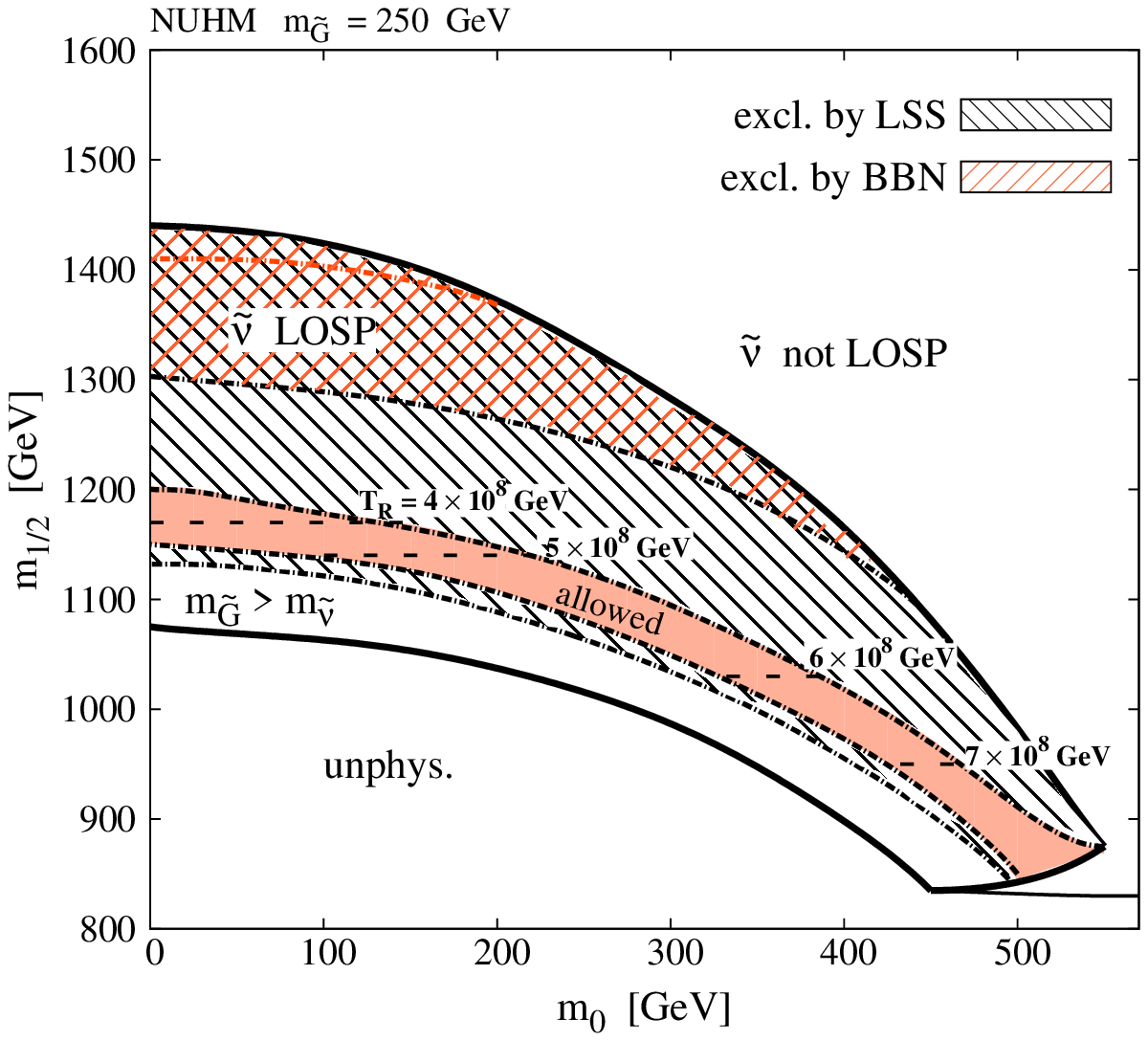}
\end{center}
\caption{A comparison of the BBN and LSS bounds for $m_{\tilde G}=175\textrm{ and }250\,\mathrm{GeV}$. Long-dashed
lines show contours of constant reheating temperature $T_R$. For ${}^6\mathrm{Li}/^7\mathrm{Li}$
the stringent limit was used;
 the boundary of the excluded region with the more
conservative constraint for ${}^6\mathrm{Li}/{}^7\mathrm{Li}$ is represented by a red dash-dotted line. 
 \label{fb3b}}
\end{figure}

As we have seen, increasing $m_{\tilde G}$ for a given $m_{\tilde
  \nu}$ tends to alleviate the BBN constraints. However, for large
gravitino masses there is another factor that we have to take into
account. Non-thermal gravitinos produced in sneutrino LOSP decays will
have velocities much larger than those characteristic for thermal
distribution. Such fast moving dark matter particles tend to erase
small scales of Large Scale Structures (LSS), especially when they constitute a
sizable fraction of the dark matter density. Following
\cite{Jedamzik:2005sx}, we account for these LSS  constraints by requiring
that the root mean square velocity of the non-thermally
produced dark matter gravitinos does not exceed 1~km/s and that the
non-thermal component makes less than 20\% of the total dark matter
abundance.

The impact of this bound on the NUHM parameter space is shown in Figure \ref{fb3b}, where we
show the superposition of the BBN bound discussed previously and the
above LSS bounds for $m_{\tilde G}=175\textrm{
  and }250\,\mathrm{GeV}$. At such large $m_{\tilde G}$, the LSS
bounds become more stringent than the BBN ones (at $m_{\tilde\nu}\geq 300\,\mathrm{GeV}$ the LOSP relic abundance, $\Omega_\mathrm{LOSP}h^2$, exceeds
20\% of the total dark matter abundance, hence $\Omega_{\tilde G}^\mathrm{NTP}h^2$
is also of this order), 
leaving just a small
allowed strip in the parameter space. For $m_{\tilde
  G}>270\,\mathrm{GeV}$, we find that the LSS bounds exclude the entire
section of the parameter space that we analyze here.  This has
important consequences for the maximum reheating temperature, since
limits on the maximum reheating temperature become weaker with increasing
gravitino mass.  

\begin{figure}
\begin{center}
\includegraphics*[height=8.8cm, trim= 0mm 0mm 0mm 0mm]{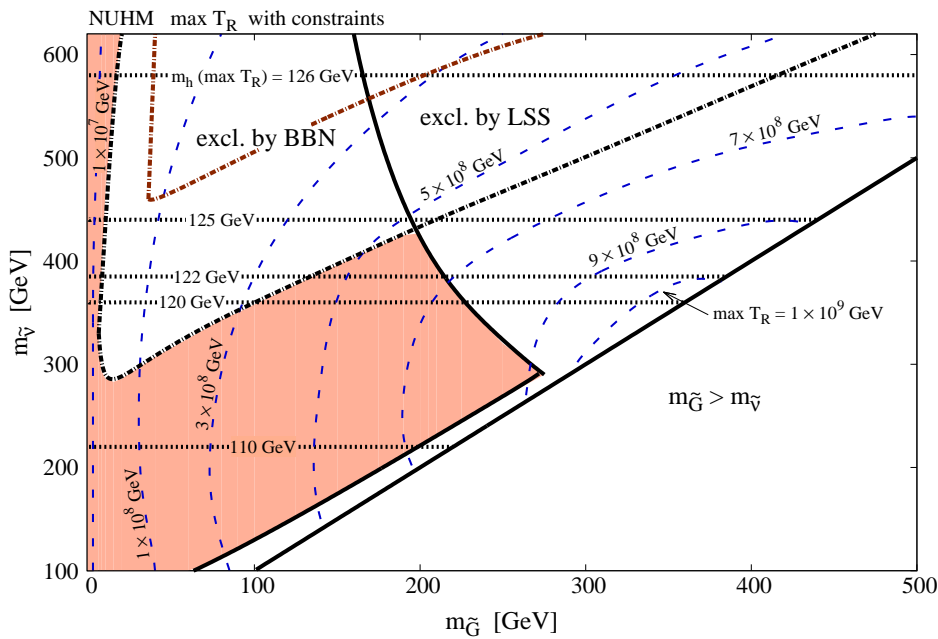}
\end{center}
\caption{A summary of the bounds in the NUHM  in the $(m_{\tilde G},m_{\tilde \nu})$ plane. 
The thick dashed line bounds the region excluded by BBN, the solid red line marks the boundary of the region excluded by LSS.
Thinner dashed lines show the maximum reheating temperature, $T_R^{\mathrm{max}}$, and thinner dotted lines show the Higgs boson mass corresponding
to $T_R^{\mathrm{max}}$.
For ${}^6\mathrm{Li}/^7\mathrm{Li}$
the stringent limit was used;
 the boundary of the excluded region with the more
conservative constraint for ${}^6\mathrm{Li}/{}^7\mathrm{Li}$ is represented by a red dash-dotted line. 
 \label{fb3c}}
\end{figure}

A summary of our results is presented in Figure \ref{fb3c} which shows
regions in the $(m_{\tilde G},m_{\tilde \nu})$ plane excluded by our
constraints.  It is clear that the BBN bounds alone allow two distinct
regions in the parameter space. For small $m_{\tilde
  G}<10\,\mathrm{GeV}$, there are no constraints on $m_{\tilde \nu}$
but the allowed maximum reheating temperature is relatively low,
$T_R^\mathrm{max}\sim10^7\,\mathrm{GeV}$. For larger $m_{\tilde G}$,
the BBN bounds start constraining the sneutrino mass and the maximum
reheating temperature increases to $\sim 10^9\,\mathrm{GeV}$ when
$m_{\tilde \nu}\sim m_{\tilde G}$. Imposing the LSS bounds closes this
second region, thus slightly reducing the reheating temperature down
to $\sim 9\times10^8\,\mathrm{GeV}$. However, now the points for which
$T_R$ is maximal correspond to Higgs boson masses much smaller than
the LHC measurement.  The requirement that the Higgs boson mass is at
least 122 GeV, brings $T_R^{\mathrm{max}}$ down to $7\times
10^8\,\mathrm{GeV}$.
 
%%%%%%%%%%%%%%%%%%%%%%%%%

\begin{figure}
\begin{center}
\includegraphics*[height=7cm, width=7cm, trim= 0mm 0mm 0mm 0mm]{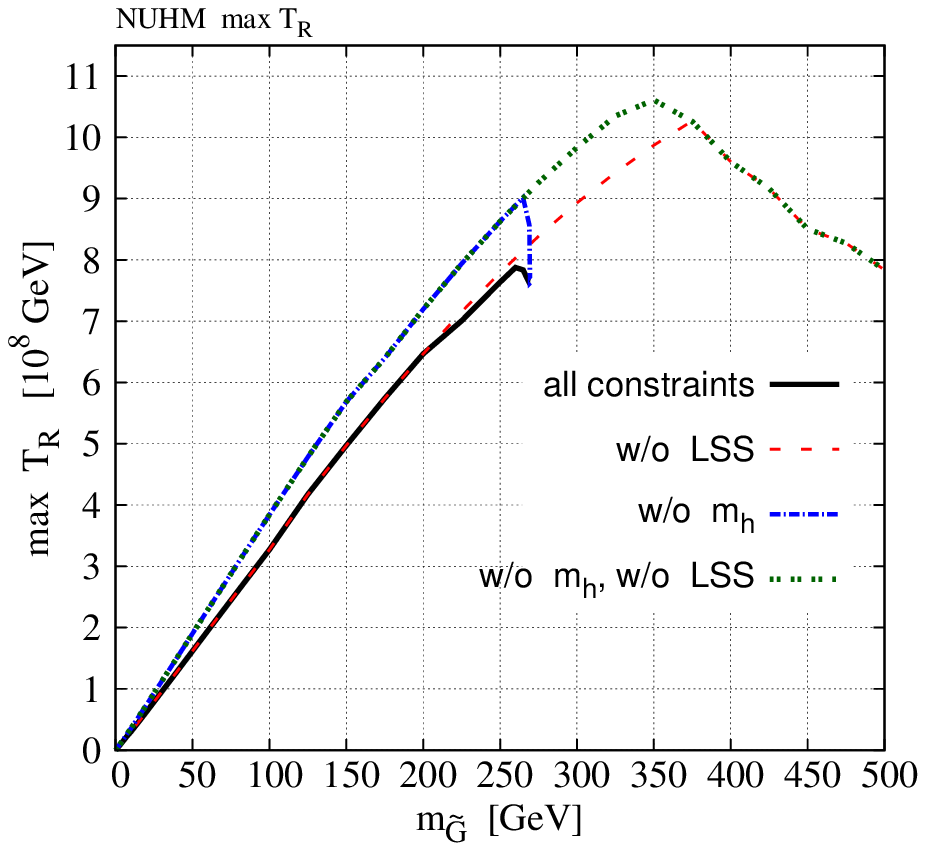}
\hspace{0.5cm}
\includegraphics*[height=7cm, width=7cm, trim= 0mm 0mm 0mm 0mm]{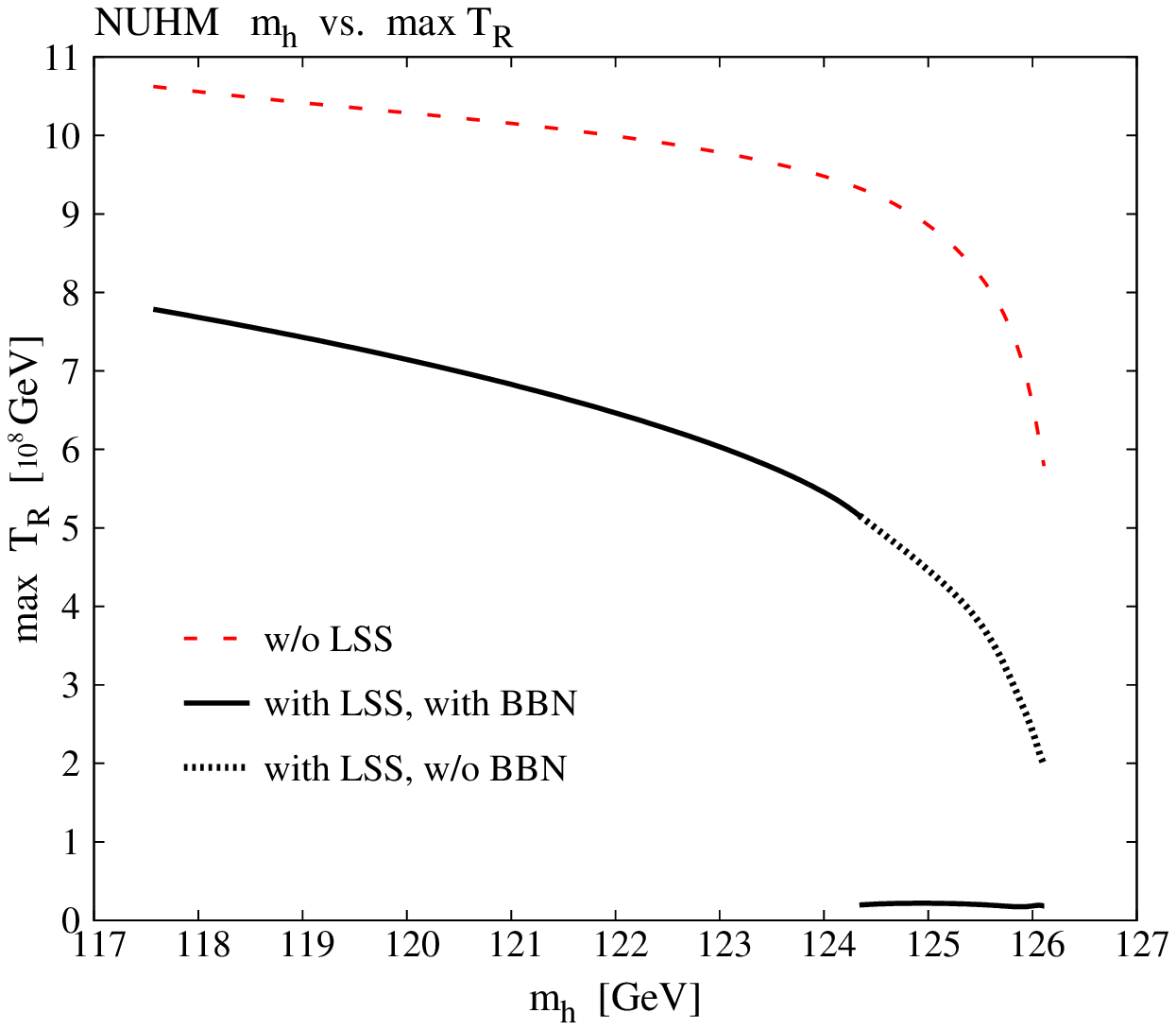}
\end{center}
\caption{ Left panel: the maximum reheating temperature as a function
  of $m_{\tilde G}$ with the BBN, the LSS and the Higgs boson mass
  constraints ($m_h>122\,\mathrm{GeV}$) applied, as well as without
  one or both of the LSS and the Higgs boson mass constraints. Right
  panel: maximum reheating temperature versus the Higgs boson mass
  without the LSS constraint (upper red dashed line) and with the LSS
  constraint (lower line). 
The solid (dotted) segments correspond
  to the cases where the BBN bound is (is not) applied. 
  \label{fb3aa}}
\end{figure}

These bounds on maximum $T_R$ as a function of $m_{\tilde G}$ are
shown in the left panel of Figure \ref{fb3aa} for the same sets of
constraints.  In the panel we impose the BBN bounds and we show the
results with and without the LSS bounds and with and without the
requirement that the Higgs boson mass is at least 122~GeV. We see that
in each case the maximum $T_R$ lies close to $10^9\,\mathrm{GeV}$,
depending on the set of bounds imposed. Without the LSS or the Higgs
boson mass bounds, this constraint mainly results from the lower bound
on $m_{1/2}$, as the maximum $T_R$ scales roughly as
$m_{1/2}^{-2}$. This can be seen in the right panel of Figure
\ref{fb3aa} where we show the maximum $T_R$ versus the Higgs boson
mass with and without BBN and LSS constraints. Note that at
$m_h=126\,\mathrm{GeV}$ the maximum $T_R$ plunges down as the BBN and
the LSS bound become inconsistent with larger values of the Higgs boson
mass.

%%%%%%%%%%%%%%%%%%%%%%%%%

These maximum values of $T_R$ is close to the quoted above lower bound required by
simple thermal leptogenesis.  It should be noted that the quoted
leptogenesis bounds should be treated as indicative rather than
absolute, since a rather mild mass degeneracy in the right-handed
neutrino sector may lower the minimum reheating temperature for
successful leptogenesis by a factor of a few \cite{Davidson:2008bu}.

%We should also mention that the BBN constraints we obtain justify {\em a posteriori} that the non-thermal contribution to DM gravitino  abundance resulting
%from decays of sneutrino LOSP can be neglected. This non-thermal contribution can be expressed as $\Omega_{\tilde G}^\mathrm{NTP}h^2=(m_{\tilde G}/m_{\tilde\nu})\Omega_{\tilde\nu}h^2$; for sneutrino masses up to 300 GeV, for which there are no BBN constraints,
%we have $\Omega_\mathrm{LOSP}h^2\sim\mathcal{O}(10^{-2})$. This LOSP relic density can be of the order of 0.1 for $m_{\tilde\nu}\sim 600\,\mathrm{GeV}$, but then BBN bounds require that $m_{\tilde G}/m_{\tilde\nu}\sim 10^{-2}$.

 \subsection{GGM models}

\begin{figure}
\begin{center}
\includegraphics*[height=7cm, width=7cm, trim= 0mm 0mm 0mm 0mm]{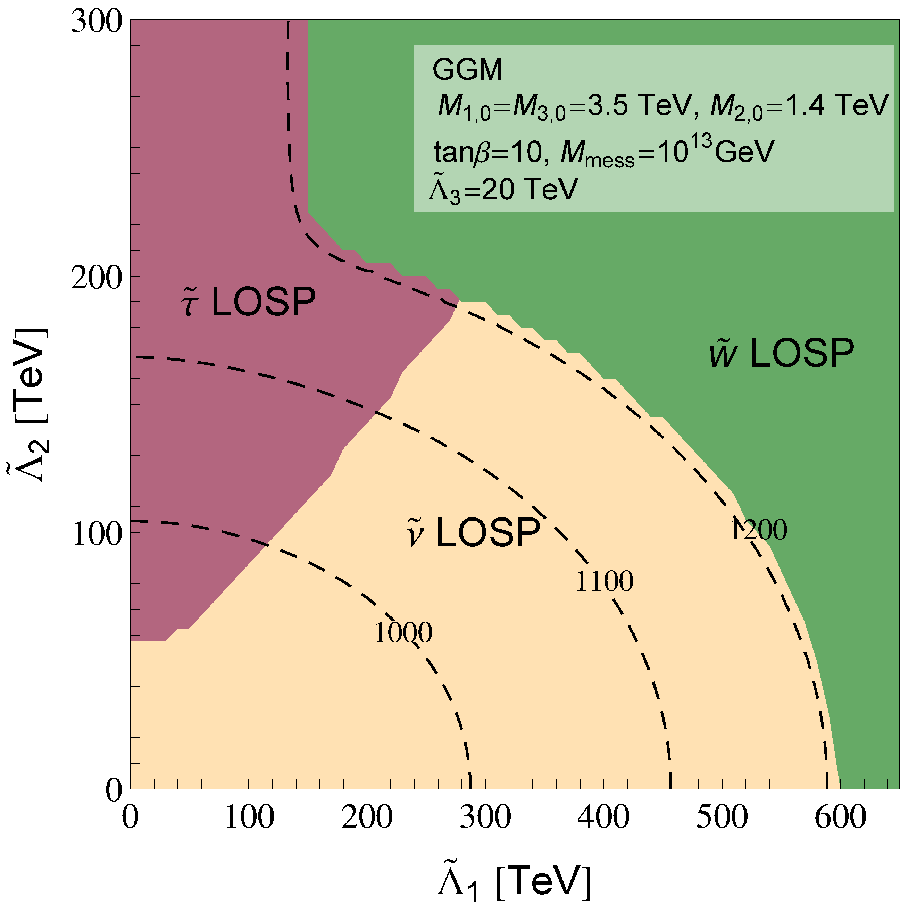}
\hspace{0.5cm}
\includegraphics*[height=7cm, width=7cm, trim= 0mm 0mm 0mm 0mm]{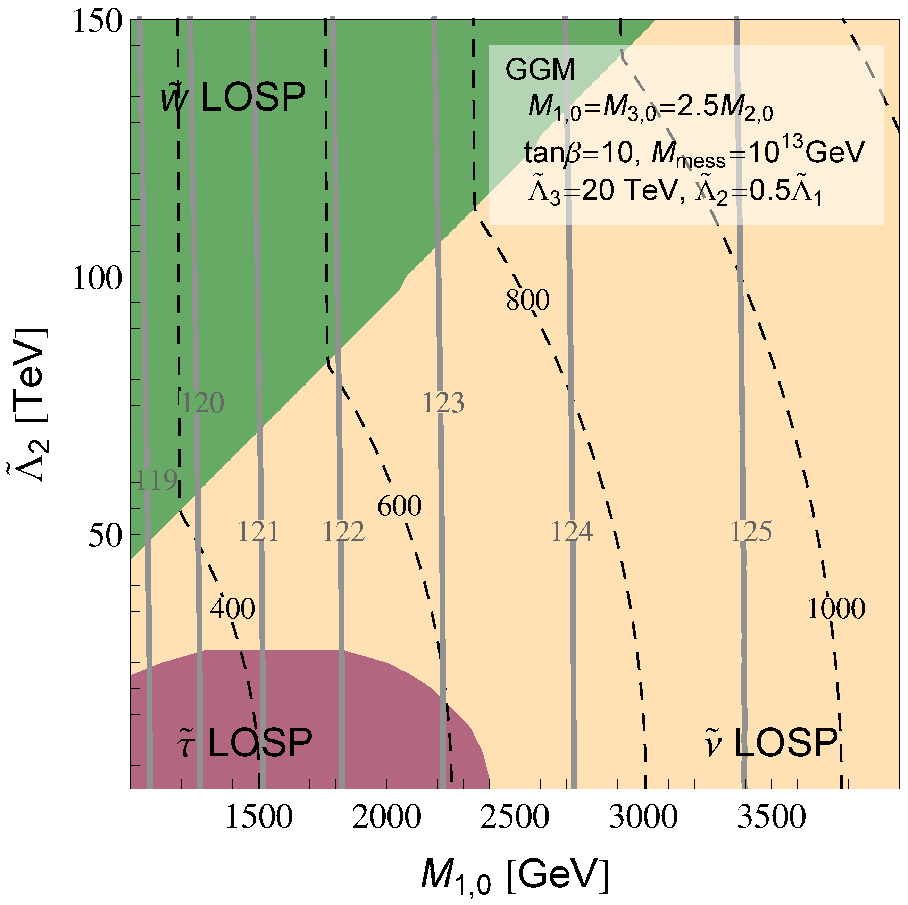}
\end{center}
\caption{
Sections of the GGM parameter space: $\tilde \Lambda_1$ vs $\tilde \Lambda_2$ (left panel) and $M_{1,0}$ vs $\tilde\Lambda_2$ (right panel) with fixed 
ratio $M_{1,0}:M_{2,0}:M_{3,0}=5:2:5$ and fixed values of $\tan\beta=10$, the messenger scale $M_\mathrm{mess}=10^{13}\,\mathrm{GeV}$
and $\tilde\Lambda_3=20\,\mathrm{TeV}$ with $\mu>0$. 
Contours of constant LOSP (Higgs boson) masses are shown as dashed (solid) lines. 
 \label{fg}}
\end{figure}

Another class of theoretically motivated scenarios in which it is possible to obtain sneutrino LOSP are models of
Generalized Gauge Mediation. Unlike in the NUHM, where the condition $\mathrm{tr}(Y\mathbf{M}_\mathrm{scalars}^2)=0$
is violated, the feature of GGM models that allows for a sneutrino LOSP is a non-universality of the gaugino masses.
In particular, it follows from (\ref{eq:solrge}) that sneutrino LOSP is viable for $M_2/M_1\simlt 2$ at the electroweak scale.
We shall therefore utilize the freedom of gaugino mass assignment offered by GGM models to 
reduce $M_{2,0}$ at the messenger scale without breaking the universality of the two remaining gauginos,
i.e.~we shall adopt $M_{1,0}=M_{3,0}$. More specifically, we shall assume $M_{1,0}:M_{2,0}:M_{3,0}=5:2:5$, which
predicts that the lightest gaugino-like neutralino is a wino.
Eqs.~(\ref{eq:solrge}) also show that for sneutrino LOSP $m_L^2$ cannot be too large,
which for fixed gaugino mass scale places upper bounds on the parameters $\tilde \Lambda_1$ and $\tilde \Lambda_2$, whose relation to scalar masses at the messenger scale is shown in the Appendix. These bounds can be seen in the left panel of Figure \ref{fg}, which also shows that increasing $\tilde\Lambda_2$ with fixed $\tilde \Lambda_1$
increases $m_{L,0}^2$ with respect to $m_{E,0}^2$, which may lead to a right-handed stau LOSP. 

The interplay between the gaugino and scalar mass scales is shown in the right panel of Figure \ref{fg}, where we keep $\tilde\Lambda_1=2\tilde\Lambda_2$. Increasing gaugino masses while keeping the mass ratios fixed enlarges the range of $\tilde\Lambda_2$
(and $\tilde \Lambda_1$) for which one of the sleptons is the LOSP. With small values of $\tilde\Lambda_1$ slepton masses are governed by 1-loop corrections proportional to gaugino masses and, as follows from (\ref{eq:solrge}), right-handed staus are the lightest. By increasing $\tilde\Lambda$
one can obtain sneutrino LOSP, because the slepton masses at the messenger scale, $m_{L,0}^2$ and $m_{E,0}^2$, contain contributions
proportional to the product of $\tilde\Lambda_1^2$ and the square of the respective hypercharge, which is larger for right-handed
sleptons.

GGM models do not allow $A$-terms at the messenger scale. (This can,
however, be circumvented by adding direct messenger-matter couplings
in the superpotential \cite{Chacko:2001km}, consistently with a Higgs
mass of 126 GeV \cite{Jelinski:2011xe}.) Therefore, in order to have a
sufficiently large Higgs boson mass, we have to consider much larger
gluino mass $M_{3,0}$ at the messenger scale, which generates
radiatively large stop masses and a large negatively $A_t$ at the
electroweak scale.  The left panel of Figure \ref{fg} corresponds to
fixed gaugino mass parameters and, therefore, to an almost constant
Higgs boson mass of 126 GeV; in the right panel, the Higgs boson mass
becomes close to 126 GeV for large values of $M_{1,0}=M_{3,0}$.  At
first sight it might seem that one could use the non-universality of
the gaugino masses to make $M_{1,0}$ and $M_{2,0}$ much smaller than
$M_{3,0}$, but a large $M_{3,0}$ also results in a large $\mu$
parameter at the electroweak scale, which prevents the sneutrino from
being the LOSP. This is also the reason for adopting a relatively
small value of $\tilde\Lambda_3$: too large squark masses at the
messenger scale also increase $\mu$. All this results in sneutrino
LOSP masses of about 1 TeV, which is much larger than in the case of
the NUHM.

\begin{figure}
\begin{center}
\includegraphics*[height=7cm, width=7cm, trim= 0mm 0mm 0mm 0mm]{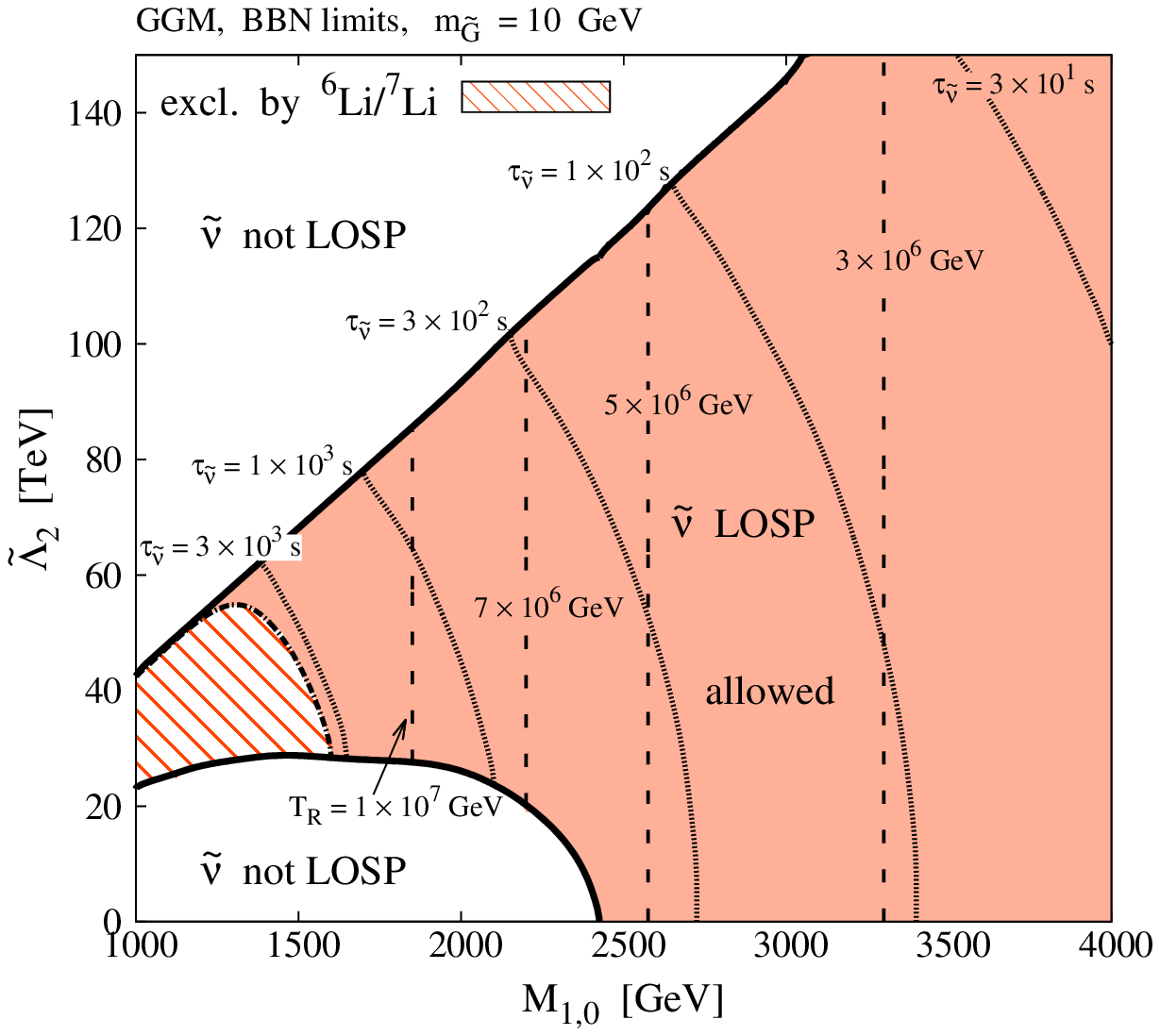}
\hspace{0.5cm}
\includegraphics*[height=7cm, width=7cm, trim= 0mm 0mm 0mm 0mm]{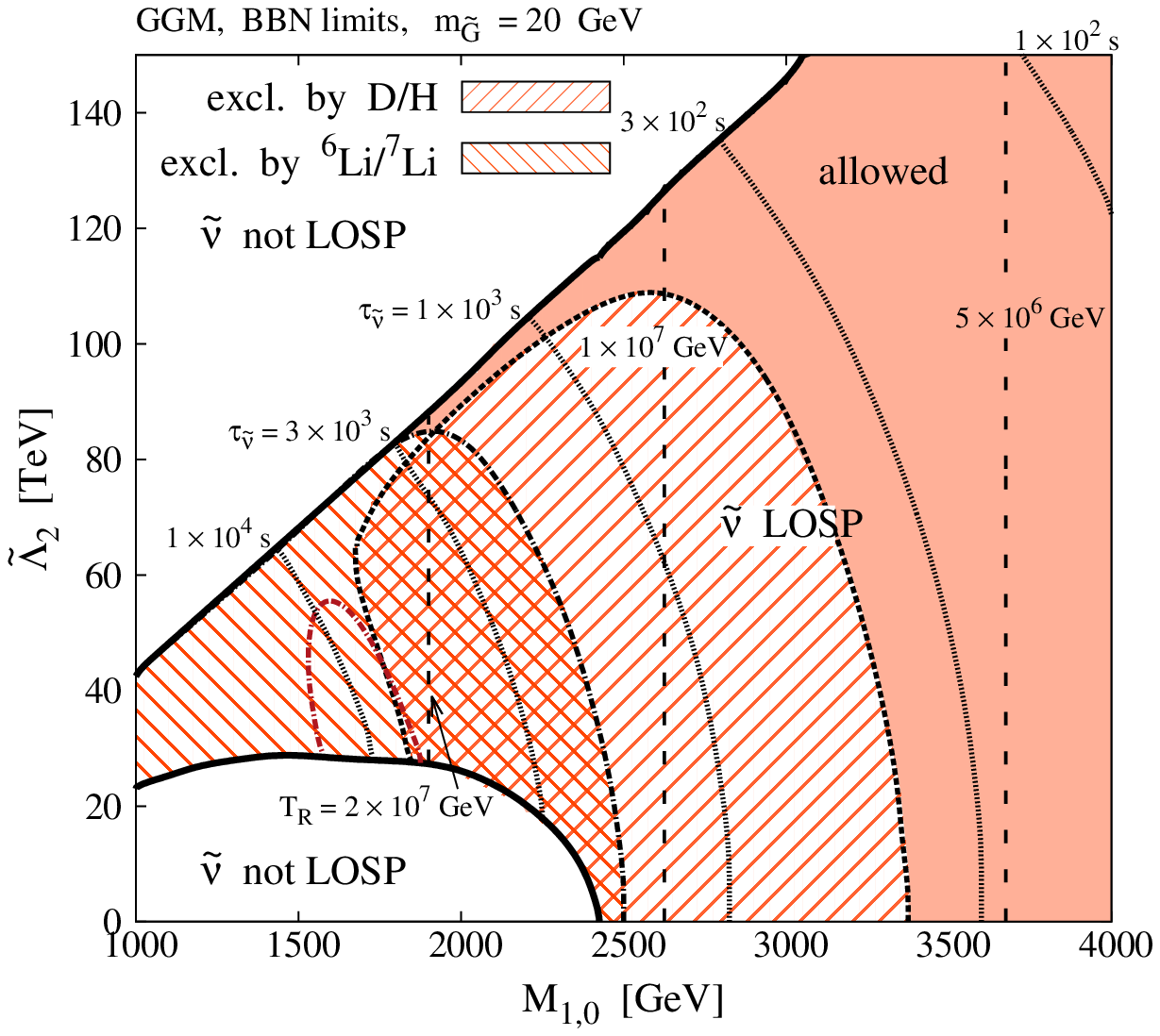}
\end{center}
\caption{
BBN bounds for the sneutrino LOSP region in GGM model shown in the right panel of Figure \ref{fg}
for values of gravitino mass $m_{\tilde G}=10\textrm{ and }20\,\mathrm{GeV}$.
For ${}^6\mathrm{Li}/^7\mathrm{Li}$
the stringent limit was used;
 the boundary of the excluded region with the more
conservative constraint for ${}^6\mathrm{Li}/{}^7\mathrm{Li}$ is represented by a red dash-dotted line. 
\label{fbg1}}
\end{figure}

Models with gauge mediation of supersymmetry breaking have the
advantage that the leading contributions to the soft masses are
flavor-diagonal, while the subdominant gravity-mediated contributions,
of the order of $m_{\tilde G}$, do not have to exhibit any such
structure. This leads to a natural suppression of the FCNC's, but
also has the obvious consequence that $m_{\tilde G}\ll m_{\tilde
  \nu}$, with the precise hierarchy depending on details of an
appropriate flavor model. For this reason we do not consider
$m_{\tilde G}$ larger than 20 GeV, hence the reheating temperature
consistent with the measured dark matter abundance is much lower for
GGM than for the NUHM. Nonetheless, we find some BBN constraints for
$m_{\tilde G}\sim\mathcal{O}(10)\,\mathrm{GeV}$; they are shown in
Figure \ref{fbg1}.

\section{Conclusions}
 
In this paper, we have analyzed models of gravitino dark matter with
the $\tau$-sneutrino as the lightest ordinary supersymmetric particle.
We have shown that if the scale at which supersymmetry is broken is
close to the unification scale, the sneutrino can be the LOSP either if gauginos
are non-universal or $D^2<0$  at the high scale. We have then performed a
detailed study of representative examples of these two possibilities:
one arising in the NUHM  and the other in models of generalized
gauge mediation of supersymmetry breaking.

We have calculated the changes in the BBN predictions due to hadronic
showers from sneutrino decays $\tilde\nu\to\nu\tilde Gq\bar q$,
calculating the hadronic energy release by a numerical integration
over the phase space of the produced particles. In the cases in which the
D/H bound  provided the strongest constraint, we observed big changes of the excluded
regions of the parameter space between this calculation and one using
a simplified formula for the hadronic energy release.

We showed that in models of gravitino dark matter with sneutrino LOSP, the four classes of constraints that can be applied:
(i) the BBN constrains, (ii) the constraints on the large structure formation due to a presence of free-streaming decay products of $\tilde{\nu}$,
(iii) the Higgs boson mass bounds derived from the LHC data and (iv) the bounds on reheating temperature
required by simple models of thermal leptogenesis are inconsistent, albeit the maximum reheating temperature is only $2-3$ times
smaller than the value suggested by the leptogenesis bound. (This was clearly visible in the NUHM; with gauge mediation,
such large reheating temperatures were unattainable due to theoretical constraints on the gravitino mass.) Therefore, our results challenge
the notion that models of gravitino dark matter with sneutrino LOSP are compatible with simple thermal leptogenesis.

%\vspace{0.3cm}
%\noindent {\bf Acknowledgements}.  
\acknowledgments This work has been funded in part
by the Welcome Programme of the Foundation for Polish Science.  LR is
also supported in part by the Polish National Science Centre grant N
N202 167440, an STFC consortium grant of Lancaster, Manchester and
Sheffield Universities and by the EC 6th Framework Programme
MRTN-CT-2006-035505.  KT is partly supported by the MNiSW grants N
N202 091839 and IP2011 056971.

%\section*{Appendix}
\appendix
\section{Soft supersymmetry breaking parameters at the high scale}
 
 Here we collect expressions for the soft supersymmetry breaking parameters at the high scale
 in the NUHM and the GGM model. 
 
 The parametrization of the NUHM is very simple
 \begin{eqnarray}
 M_{1,0}&=&M_{2,0}=M_{3,0}=m_{1/2} \\
 m_{Q,0}^2&=&m_{U,0}^2=m_{D,0}^2=m_{L,0}^2=m_{E,0}^2=m_0^2 \, ,
 \end{eqnarray}
 while $m_{H_u}^2$, $m_{H_d}$, $A_0$, $\tan\beta$ and $\mathrm{sgn}(\mu)$ can are free
 parameters.
 
In GGM models, the soft supersymmetry breaking masses at the high scale in the notation of \cite{Carpenter:2008wi} read
\begin{eqnarray}
M_{r,0} &=& (\alpha_r/4\pi)\Lambda_1 \qquad\textrm{for}\,\,r=1,2,3 \\
m_{Q,0}^2 &=& (8/3)(\alpha_3^2/16\pi^2)\tilde\Lambda_3^2+(3/2)(\alpha_2^2/16\pi^2)\tilde\Lambda_2^2+(1/30)(\alpha_1^2/16\pi^2)\tilde\Lambda_1^2 \\
m_{U,0}^2 &=& (8/3)(\alpha_3^2/16\pi^2)\tilde\Lambda_3^2+(8/15)(\alpha_1^2/16\pi^2)\tilde\Lambda_1^2 \\
m_{D,0}^2 &=& (8/3)(\alpha_3^2/16\pi^2)\tilde\Lambda_3^2+(2/15)(\alpha_1^2/16\pi^2)\tilde\Lambda_1^2 \\
m_{L,0}^2 &=& m_{H_u}^2 = m_{H_d}^2 = (3/2)(\alpha_2^2/16\pi^2)\tilde\Lambda_2^2+(3/10)(\alpha_1^2/16\pi^2)\tilde\Lambda_1^2 \\
m_{E,0}^2 &=&(6/5)(\alpha_1^2/16\pi^2)\tilde\Lambda_1^2 \, . 
\end{eqnarray}
The trilinear scalar couplings are all equal to zero and $\tan\beta$, $\mathrm{sgn}(\mu)$ are free parameters.

\end{document}